\documentclass[graybox, envcountchap]{svmult}

\usepackage{amsmath}
\usepackage{amssymb}
\usepackage{color}
\usepackage{helvet}          
\usepackage{courier}         
\usepackage{dirtree}

\usepackage{makeidx}        
\usepackage{graphicx}   
\usepackage{overpic}
\usepackage{epstopdf}
\usepackage{epsfig}

\usepackage{multicol}        
\usepackage[bottom]{footmisc}

\usepackage[misc]{ifsym}

\newcommand{\bea}{\begin{eqnarray}}
\newcommand{\eea}{\end{eqnarray}}
\newcommand{\be}{\begin{equation}}
\newcommand{\ee}{\end{equation}}

\makeindex             

\begin{document}


\title{Semi-classical dust collapse and regular black holes}
\author{Daniele Malafarina}
\institute{Daniele Malafarina (\Letter) \at Department of Physics, Nazarbayev University, Kabanbay Batyr 53, 
\\010000 Astana (Kazakhstan),\\ \email{daniele.malafarina@nu.edu.kz}}
%
%
\maketitle

\abstract{Semi-classical corrections at large curvature are employed in toy models of spherically symmetric gravitational collapse in order to avoid the formation of singularities. The resulting spacetimes may produce bounces, compact remnants or regular black holes in place of the usual Schwarzschild black hole. Within these models, a whole class of collapse scenarios leading to the formation of regular black holes may be obtained from General Relativity coupled to some theory of non-linear electrodynamics. In the present chapter we provide a thorough exposition of semi-classical dust collapse with particular attention to the conditions for the formation of regular black holes as the endstate of collapse.}


\section{Introduction}
\label{sec:1}


In 1939 Oppenheimer and Snyder \cite{OS} and independently Datt \cite{Datt} developed the first mathematical model describing complete gravitational collapse in General Relativity (GR). The model, usually referred to as OSD, is given by an exact solution of the field equations for a dynamical spherically symmetric cloud of homogeneous collisionless matter, usually called `dust', collapsing under its own gravity.
Homogeneous dust collapse results in the formation of a Schwarzschild black hole where the central spacetime singularity, that develops as the endstate of collapse, is covered by the horizon at all times.

We now know, from the singularity theorems, that singularities must inevitably appear as the endstate of collapse once a series of conditions are met. These conditions are (i) the validity of GR during collapse, (ii) the validity of some energy condition, (iii) global hyperbolicity of the spacetime and (iv) the formation of trapped surfaces at some point during collapse. Taken as hypotheses, these assumption lead to a series of theorems that show that singularities are an inevitable outcome of collapse \cite{sing3,sing1,sing,sing2}. 

One could conjecture that singularities must always be hidden behind horizons in order to preserve the causal structure of the spacetime \cite{ccc}. However, even before the formulation of the singularity theorems, many researchers speculated that singularities should not form in the real universe and therefore one or more of the above hypotheses must be violated at some stage during collapse. Intuitively speaking this means that some kind of repulsive effect must appear to halt the attractive force of gravity before the formation of the singularity.
We know that for objects that are sufficiently massive and sufficiently compact the known forces of nature are not able to halt collapse \cite{OV}. Therefore the repulsive effects must come from some physics that is yet unknown to us and dominates when the spacetime curvature becomes large enough. This may be due to the effects of a new theory of gravity, thus modifying hypothesis (i) \cite{collapse1}, and/or to a modification of the other forces and thus the averaged properties of matter, thus modifying hypothesis (ii)\cite{Bergmann}.

In order to investigate the implications of such modifications for black hole physics several toy models have been developed over the past few decades. These include modifications of black hole geometries such as \cite{bh6,bh1,bh2,bh9,bh10,bh7,bh5,bh3,bh4,bh8} as well as modified collapse models such as \cite{collapse6,collapse5,collapse7,collapse2,collapse3,collapse4,collapse9,collapse8}. For a recent review see \cite{universe}. In turn, these modifications often present interesting features that may bear significant consequences for astrophysical black hole candidates. 
One of the most interesting class of modified black hole geometries is that of regular black hole solutions obtained within some theory of non linear electrodynamics (NLED) \cite{nled1,nled6,nled2,nled3,nled7,nled4,nled5}.
Then one is naturally led to consider under what circumstances such NLED regular black holes can develop from collapse \cite{bobir}. In the present chapter we provide a detailed construction of semi-classical models for dust collapse and investigate the conditions under which they may lead to black holes, bounces or regular black holes as final states.

The chapter is organized as follows: In section \ref{sec1} we review the field equations for spherical collapse, the most important features of collapse solutions and the general procedure to formulate semi-classical models. Section \ref{sec2} is devoted to the spacetimes describing the exterior of the collapsing sphere with particular attention to regular black holes in NLED. In section \ref{sec3} we review the formalism for matching the interior and exterior geometries across a collapsing time-like surface, while in section \ref{sec4} we discuss in detail dust collapse with semi-classical corrections and the conditions for the formation of a regular black hole.
Finally section \ref{sec5} provides a brief summary and conclusions.

Throughout the chapter we shall adopt the convention of absorbing the factor $8\pi$ in Einstein's equations into the definition of the energy-momentum tensor and we will use geometrized units taking $G=c=1$.

\section{Interior: Gravitational collapse}\label{sec1}

Let us first review the general formalism to describe relativistic collapse \cite{joshi-book}. We shall consider the line-element for the spherical collapsing interior in co-moving coordinates $\{t,r,\theta,\phi\}$ as \cite{review, chapter}
\be \label{line}
ds^2=-e^{2\nu}dt^2+e^{2\psi}dr^2+C^2d\Omega^2\;,
\ee 
where $d\Omega^2=d\theta^2+\sin^2\theta d\phi^2$ is the line element on the unit 2-sphere and the metric functions depend only on $t$ and $r$, namely $\nu(t,r)$, $\psi(t,r)$, $C(t,r)$. Notice that $r$ is a coordinate radius `attached' to the infalling particles while $C$ is the so-called area-radius function related to the area of collapsing spheres of constant $r$. Then collapse is described by $C$ decreasing in time.
For the energy momentum tensor we shall start with an anisotropic inhomogeneous fluid with
\be 
T^{\mu\nu}=(\varepsilon+p_\theta)u^\mu u^\nu+ p_\theta g^{\mu\nu}+(p_r-p_\theta)\xi^\mu \xi^\nu\;,
\ee 
where $\varepsilon (t,r)$ is the energy density, $p_r (t,r)$ and $p_\theta (t,r)$ are the radial and tangential pressures, $u^\mu$ is the fluid's four velocity and $\xi^\mu$ is a space-like unit vector orthogonal to $u^\alpha$.
We can define a mass function $F$ known as Misner-Sharp mass \cite{misner} as
\be \label{misner}
F=C(1-e^{-2\psi}C'^2+e^{-2\nu}\dot{C}^2)=C(1-G+H) \;,
\ee 
where we used primed quantities for partial derivatives with respect to $r$ and dotted quantities for derivatives with respect to $t$ and we have introduced two new functions defined as
\be 
G=e^{-2\psi}C'^2, \;\; H=e^{-2\nu}\dot{C}^2\;.
\ee 
The Misner-Sharp mass may intuitively being understood as describing the amount of matter contained within the shell $r$ at the time $t$.
Additionally we need to consider conservation of energy momentum, i.e. the Bianchi identities, given by $\nabla_\mu T^{\mu\nu}=0$ which for the zero component gives
\be \label{Bianchi}
\nu'=-\frac{p_r '}{\varepsilon+p_r}+2\frac{p_\theta-p_r}{\varepsilon+p_r}\frac{C'}{C}\;.
\ee 
In the case of a perfect isotropic fluid, given by $p_\theta=p_r=p$, the Bianchi identity becomes
\be \label{bianchi}
\nu'=-\frac{p'}{\epsilon+p}\;,
\ee 
and the field equations then become
\bea \label{EE-rho}
\varepsilon&=&\frac{F'}{C^2C'}\;, \\ \label{EE-p}
p_r&=&p=-\frac{\dot{F}}{C^2\dot{C}}\;, \\ \label{EE-G}
2\dot{C}'&=&C'\frac{\dot{G}}{G}+\dot{C}\frac{H'}{H}\;.
\eea 
using equation \eqref{bianchi} we can also rewrite equation \eqref{EE-G} as
\be
\frac{\dot{G}}{G}=2\nu'\frac{\dot{C}}{C'}\;.
\ee 
We are left with a set of five equations, namely three field equations, one Bianchi identity and the definition of the Misner-Sharp mass, for six unknown quantities, i.e.  $F$, $p$, $\epsilon$ and three metric functions $\nu$, $\psi$ and $C$. Therefore in order to close the system we need one additional relation. This is usually given in the form of an equation of state relating density and pressure $p=p(\epsilon)$.
In the following we will consider the simplest case of non interacting particles, also called `dust', for which $p=0$.
However, for completeness, it is worth mentioning the most commonly used equations of state in astrophysics and cosmology, which are the linear barotropic equation
\be 
p=\omega\epsilon\;,
\ee
with $\omega\in[-1,1]$ and the polytropic equation
\be 
p=K\epsilon^{(n+1)/n}\;,
\ee 
with the polytropic index $n$ usually taken between $0.5$ and $1$ for compact objects such as neutron stars \cite{chandra-book,Tooper}.

\subsection{Regularity and scaling}

To ensure that the density and pressures are regular and the mass function $F$ is well behaved at the center at the initial time $t_i$ some additional conditions are necessary\cite{JD}. In fact it is immediately clear from equation \eqref{EE-rho} that there exist mass functions $F$ for which the density diverges at $C=0$ at all times. If we wish for collapse to start from a regular configuration we must restrict the allowed functions $F$ to those that give a finite density everywhere at $t_i$. Since we still have some gauge freedom in specifying the initial value of the area-radius function $C$ we can impose
\be 
C(t_i,r)=r\;,
\ee 
and thus we see that we can define an adimensional function $a(t,r)$, usually called `scale factor', in such a way that
\be \label{scaling}
C(t,r)=ra(t,r)\;,
\ee 
with the initial condition for $a$ set as
\be \label{initial-a}
a(t_i,r)=1\;.
\ee 
Then the condition for collapse is given by $\dot{a}<0$ and collapse ends in a singularity if $a\rightarrow 0$ in a finite time. Using the scaling \eqref{scaling} and equation \eqref{initial-a} in the field equation for the energy density \eqref{EE-rho} we get the initial condition for the density as
\be 
\varepsilon(t_i,r)=\frac{F'(t_i,r)}{r^2(1+ra'(t_i,r))}\;,
\ee 
which diverges for $r\rightarrow 0$ unless $F\sim r^n$ with $n\geq 3$ for $r$ close to zero. Then we are led to impose the following rescaling for the Misner-Sharp mass
\be 
F(t,r)=r^3m(t,r)\;,
\ee 
with $m(t_i,0)={\rm const}.\neq 0$, which ensures that the mass function is sufficiently regular at the center at the initial time and the singularity may develop only at a later time. Also, from the definition of the Misner-Sharp mass \eqref{misner} we see that the introduced scaling leads to
\be \label{misner-2}
m=a\left(\frac{1-G}{r^2}+e^{-2\nu}\dot{a}^2\right)\;.
\ee 
Imposing regularity of $m$ at the center then imposes the additional condition
\be 
G(t,r)=1+r^2b(t,r)\;.
\ee 
with $b(t_i,0)={\rm const}.\neq 0$.
To summarize, we have imposed the following rescaling
\bea 
C(t,r)&=&ra(t,r)\;, \\
F(t,r)&=&r^3m(t,r)\;, \\
G(t,r)&=&1+r^2b(t,r)\;,
\eea
and thus we can rewrite the field equations, the Misner-Sharp mass and the Bianchi identity for an isotropic perfect fluid as
\bea \label{EE-rho2}
\varepsilon&=&\frac{3m+rm'}{a^2(a+ra')}\;,\\ \label{EE-p2}
p&=&- \frac{\dot{m}}{a^2\dot{a}}\;,\\ \label{EE-G2}
\frac{r\dot{b}}{1+r^2b}&=&2\nu'\frac{\dot{a}}{a+ra'}\;,\\ \label{nu2}
\nu'&=&-\frac{p'}{\varepsilon+p}\;,\\ \label{misner-3}
m&=&a(e^{-2\nu}\dot{a}^2-b)\;.
\eea 
If $a\rightarrow 0$ in a finite co-moving time the density $\epsilon$ diverges and it can be shown that the Kretschmann scalar, which is the invariant scalar obtained from the Riemann tensor as $\mathcal{K}=R_{\alpha\beta\mu\nu}R^{\alpha\beta\mu\nu}$, also diverges, thus giving rise to a true curvature singularity.

Notice that if the pressure does not vanish there is one more condition to impose to ensure that the initial pressure $p_i=p(t_i,r)$ is finite. In fact if $\dot{a}(t_i,r)=0$ for some value of $r$ from equation \eqref{EE-p2} we see that $p_i$ might diverge. From equation \eqref{misner-3} we get
\be 
\frac{\dot{m}}{\dot{a}}=e^{-2\nu}\left(\dot{a}^2+2a\ddot{a}-2\dot{\nu}a\dot{a}\right)-b-\frac{\dot{b}a}{\dot{a}}\;,
\ee 
from which we see that if $b=b(r)$ then $p_i$ is finite at $t_i$ even if $\dot{a}(t_i,r)=0$ provided that $\nu$, $b$ and $a$ are well behaved. On the other hand if $b=b(t,r)$ then an additional condition on $b(t_i,r)$ must be imposed ensuring that $\dot{b}/\dot{a}$ is finite at $t_i$. Of course this condition is not necessary in the case of dust.

\subsection{Trapped surfaces, singularities and energy conditions}

Gravitational collapse produces a black hole when some kind of trapped surface appears as matter collapses from an initially non trapped configuration \cite{horizon,horizon1,horizon2,horizon3}.
For spherical collapse models we can define the apparent horizon as the curve $t_{\rm ah}(r)$ for which the surface $C(t_{\rm ah}(r),r)$ becomes null. Namely from the metric this condition can be written as
\be
X(t,r)=g^{\mu\nu}(\partial_\mu C)(\partial_\nu C)=0 \; .
\ee
Then according to the implicit function theorem $X(t,r)=0$ describes implicitly the curve $t_{ah}(r)$ (or $r_{ah}(t)$) which gives the time at which the shell $r$ becomes trapped.
Applying the definition of the Misner-Sharp mass from equation \eqref{misner} to the above equation we obtain the condition for the formation of trapped surfaces as
\be\label{trapped}
1-\frac{F}{C}=1-\frac{r^2m(t,r)}{a(t,r)}=0 \; .
\ee
Then, looking at equation \eqref{misner-3} the apparent horizon curve $t_{ah}(r)$ is given by the condition
\be\label{apparent-horizon}
\frac{r^2e^{-2\nu}}{1+r^2b}=\frac{1}{\dot{a}^2} \; .
\ee
Keep in mind that in general we may have $b=b(t,r)$ and $\nu=\nu(t,r)$, however for dust collapse, as we shall see later, we have $b=b(r)$ and $\nu=0$ making the left-hand side a function of $r$ only.

In constructing a collapse model one wants to start with a configuration that has no trapped surfaces. This can be done by imposing that the at the initial time the solutions of equation \eqref{apparent-horizon}, if any, are located outside the boundary of the cloud $r_b$, i.e. equation \eqref{apparent-horizon} has no solutions for $r\leq r_b$ at $t=t_i$, or equivalently  $r_{\rm ah}(t_i)>r_b$.

As mentioned collapse ends in a spacetime singularity if $a\rightarrow 0$. This condition can also be described via a curve $t_s(r)$ denoting the time at which the shell $r$ becomes singular.
In this case, all geodesics located inside the trapped region must terminate at the singularity. Again the curve $t_s(r)$ can be given implicitly by
\be
a(r,t_s(r))=0 \; .
\ee
Notice that in the OSD model, with the energy density being homogeneous, we get that $t_s(r)={\rm const}.$ and $t_{\rm ah}(r)$ is monotonically decreasing, which means that the singularity at the end of collapse is always covered by the trapped surface.
However, even in simple inhomogeneous collapse models such as the Lema\`itre-Tolman-Bondi dust case \cite{L,T,B} this is not obvious. In fact there exist models with $\epsilon(t_i,r)$ decreasing outwards in $r$ where both $t_s(r)$ and $t_{\rm ah}(r)$ are monotonically increasing outwards and $t_s(0)=t_{\rm ah}(0)$, thus leaving the first point of the singularity curve not necessarily covered by the trapped surface \cite{joshi}. In both cases $t_{\rm ah}(r)\leq t_s$ for all $r$ (with $t_{\rm ah}(0)=t_s(0)$) and therefore for $r>0$ the singularity is covered \cite{chapter}.

The energy momentum tensor in Einstein's equations describes the averaged properties of matter at macroscopic scales. Therefore conditions must be imposed to ensure that it describes physically viable matter fields. To this aim there are three inequalities that can be imposed for $T_{\mu\nu}$ to be considered physically valid
\cite{HE}:
\begin{enumerate}
    \item The weak energy condition (w.e.c.) states that $T_{\mu\nu}$ must satisfy the condition $T_{\mu\nu}V^\mu V^\nu\geq0$ for any time-like (and null) vector $V^\mu$. This implies that the energy density must be non negative in any reference frame. Then the weak energy conditions in the co-moving frame can be written as
\be
\epsilon\geq 0 \; , \; \epsilon+p\geq 0 \; .
\ee
\item The additional requirement that the total mass is conserved leads to the dominant energy condition (d.e.c.). This implies $T_{\mu\nu}V^\mu V^\nu\geq0$ for every time-like vector $V^\mu$ and $T_{\mu\nu}V^{\mu}$ must be null or time-like. This is a more stringent condition with respect to the w.e.c. because it also requires that the flow of $\epsilon$ must be locally non space-like. In the co-moving frame used here this translates to the additional requirement that the energy density must be greater than the pressures. Namely
\be
\epsilon\geq 0 \;, \; -\epsilon\leq p\leq \epsilon  \; .
\ee
\item Finally the strong energy condition (s.e.c.) requires that for every time-like vector $V^\mu$ we have $(T_{\mu\nu}-g_{\mu\nu}T/2)V^\mu V^\nu\geq 0$. In the co-moving frame the s.e.c. requires
\be
\epsilon\geq 0 \;, \; \epsilon+p\geq 0 \;, \; \epsilon +3p\geq 0 \;.
\ee
\end{enumerate}

Notice that the w.e.c. does not require the conservation of the baryon number of $T_{\mu\nu}$ and therefore new particles can be created if one does not impose other energy conditions. The d.o.c. is more stringent than the w.e.c. as it requires mass conservation and also it does not allow for faster than light speed of sound in the medium. Finally the s.e.c. is more stringent than the other two and it may be violated by physically valid matter models such as scalar fields.
It is important to note that the energy conditions refer to the behavior of matter at macroscopic scales and they need not apply to matter fields in the strong curvature regime, close to the formation of the singularity, where quantum effects and some corresponding quantum energy conditions may dominate \cite{visser-ec,visser-ec2}. Therefore models allowing for violations of the energy conditions towards the end of collapse may be considered even within GR.

The singularity theorems tell us that in GR if the energy conditions are satisfied and a trapped surface appears during collapse then a singularity must form. Therefore in order to avoid the formation of the singularity while retaining the formation of trapped surfaces, we must require that either GR does not hold for the whole collapse and/or that energy conditions are violated at some point.

\subsection{Semi-classical collapse}

As mentioned earlier in order to avoid the formation of a singularity at the end of collapse one or more of the hypothesis of the singularity theorems must be violated. Keeping the assumption that the spacetime be globally hyperbolic and assuming that trapped surfaces do form during collapse (after all we do observe black hole candidates in the universe) we may look for violations of the energy conditions or a breakdown of GR in the last stages of collapse.
We may then describe both scenarios in a unified formalism if we further assume that the breakdown of GR takes place in a way that can be written in the form of Einstein's equations with semi-classical corrections, namely assuming that the field equations become
\be \label{Einstein}
G_{\mu\nu}+<G_{\mu\nu}>=T_{\mu\nu}\;,
\ee 
where $T_{\mu\nu}=T_{\mu\nu}^{\rm matter}$ accounts for the matter content including the part of the energy momentum tensor violating energy conditions, if any, while $<G_{\mu\nu}>=G_{\mu\nu}^{\rm corr}$ is obtained from averaging the effects of the modifications to the geometry in such a way that the new theory still obeys Einstein equations for an effective geometry $g^{\rm eff}_{\mu\nu}=g_{\mu\nu}+<g_{\mu\nu}>$ \cite{collapse5}.
Then we can bring $G_{\mu\nu}^{\rm corr}$ on the right-hand side of equation \eqref{Einstein} and treat it as an additional, non physical, component of the energy-momentum tensor.  From the above considerations we obtain the effective energy momentum tensor as
\be 
T_{\mu\nu}^{\rm eff}=T_{\mu\nu}^{\rm matter}+T_{\mu\nu}^{\rm corr} \, ,
\ee 
with the strong field corrections to GR now described by $T_{\mu\nu}^{\rm corr}=-G_{\mu\nu}^{\rm corr}$. Keep in mind that $T_{\mu\nu}^{\rm corr}$ does not describe a matter source. Therefore even if $T_{\mu\nu}^{\rm matter}$ obeys the energy conditions we may have that $T_{\mu\nu}^{\rm eff}$ violates them as a consequence of the modifications to the theory.
We can also write the action for the semi-classical collapse model as
\be 
\mathcal{A}=\frac{1}{2}\int d^4x\sqrt{|g|}\left(\mathrm{R}+\mathcal{L}_{\rm matter}+\mathcal{L}_{\rm corr}\right) \, ,
\ee 
where $\mathcal{L}_{\rm corr}$ is the Lagrangian density describing the strong curvature corrections to GR.

In the case of a perfect fluid $T^{\mu\nu}_{\rm matter}$ is given by 
\be 
T^{\mu\nu}_{\rm matter}=(\epsilon+p) u^\mu u^\nu+pg^{\mu\nu}\;,
\ee
with $u^\kappa$ being the 4-velocity of the fluid. 
In typical scenarios we may define a critical density $\epsilon_{\rm cr}$ for which the deviations from GR become non negligible and write $T^{\mu\nu}_{\rm corr}$ as an expansion in $\epsilon/\epsilon_{\rm cr}$ close to zero, i.e. for $\epsilon<< \epsilon_{\rm cr}$. Then the effective density be written as
\be \label{eps-eff}
\epsilon_{\rm eff}=\epsilon+\alpha_1\epsilon^2+\alpha_2\epsilon^3+... \, ,
\ee 
where the parameters $\alpha_i$ depend on $\epsilon_{\rm cr}$ and are obtained from the expansion of $T^{\mu\nu}_{\rm corr}$.

In Einstein's equations the geometry side of the equations remains unchanged while the matter fields in the field equations \eqref{EE-rho2} and \eqref{EE-p2} can still be written in the same form with the effective quantities replacing the classical ones. Namely we get
\bea \label{EE-rho3}
\epsilon_{\rm eff}&=&\frac{3m_{\rm eff}+rm_{\rm eff}'}{a^2(a+ra')}\;,\\ \label{EE-p3}
p_{\rm eff}&=&- \frac{\dot{m}_{\rm eff}}{a^2\dot{a}}\;,
\eea 
with the effective Misner-Sharp mass being
\be 
m_{\rm eff}=a(e^{-2\nu}\dot{a}^2-b)\;.
\ee 

As expected, depending on the specific choice of the effective energy momentum tensor the final outcome of collapse, need not necessarily be a singularity. Besides black holes one may obtain models that bounce, models that `evaporate' and models that settle to massive compact remnants. One notable example of a bouncing model was considered in \cite{collapse6} and it is given by the choice $\alpha_1=-1/\epsilon_{\rm cr}$ and $\alpha_i=0$ for $i>1$. This case, while being the simplest possible, is also well motivated as it arises from the effective description proposed in Loop Quantum Cosmology \cite{collapse2,LQC2,LQC3}. 
The aim then is to construct a physically well motivated $T_{\mu\nu}^{\rm eff}$, solve Einstein's equations to obtain $a(t,r)$ and then investigate whether a singularity occurs and the behavior of the trapped region delimited by $t_{\rm ah}(r)$.

\section{Exterior: Regular black holes}\label{sec2}

To have a global geometry for the model we need to provide a line-element for the exterior spacetime to match to the line element \eqref{line}, which describes the collapsing interior, across a suitable boundary $r_b$. 
The natural choice to match the OSD model is a Schwarzschild exterior, which describes a static black hole once the boundary of the collapsing cloud crosses the horizon. However, modified models may require for the exterior to be modified accordingly. 
Similarly to what has been discussed for semi-classical collapse one can consider a semi-classical description of the black hole geometry. Starting from the Schwarzschild solution describing a classical static black hole one then can devise new non vacuum solutions, where the non vanishing energy momentum tensor is interpreted as the effective correction $T_{\mu\nu}^{\rm corr}$. In turn these solutions may not present a central singularity.
For example one may consider the exterior line element in coordinates $\{T, R, \theta,\phi\}$ written as
\be \label{metric-rbh}
ds^2=-f(R)dT^2+\frac{dR^2}{f(R)}+R^2d\Omega^2\;,
\ee 
where, in analogy with the Schwarzschild case, we can take $f$ defined in terms of a mass function $M(R)$ as
\be \label{F-nled}
f(R)=1-\frac{2M(R)}{R}\;,
\ee 
with $M(R)\rightarrow M_0$ for $R$ large in order to retrieve the Schwarzschild solution in the weak field where semi-classical corrections are negligible.
Obviously these are not vacuum solutions, and, as said, the energy momentum can be understood as a semi-classical correction to the Schwazrschild vacuum. The energy momentum tensor for the metric \eqref{metric-rbh} is
\bea 
T_0^0&=&T_1^1=-\frac{2 M'(R)}{R^2}\, , \\
T_2^2&=&T_3^3=-\frac{M''(R)}{R}\, .
\eea 
from which we see that one must choose $M(R)$ in such a way that $T^{\mu\nu}$ goes to zero at large distances.
The Kretschmann scalar for this spacetime is
\be \label{K}
\mathcal{K}=\frac{48M^2}{R^6}-\frac{16M}{R^3}\left(\frac{4M,_{R}}{R^2}-\frac{M,_{RR}}{R}\right)+4\left(\frac{8M,_{R}^2}{R^4}-\frac{4M,_{R}M,_{RR}}{R^3}+\frac{M,_{RR}^2}{R^2}\right),
\ee 
and the condition for avoidance of the central singularity is then given by $M(R)/R^3$ being finite for $R\rightarrow 0$.
Then noting that in this case $F(R)\rightarrow 1$ both for $R\rightarrow 0$ and $R\rightarrow +\infty$ we see that there must be either zero or an even number of roots of $F(R)=0$, corresponding to no horizons or an even number of horizons. The simplest case where horizons are present is that of two horizons, namely an outer one, corresponding to the black hole event horizon and an inner one, which is a Cauchy horizon. In fact if the dominant energy condition holds then it can be shown that the number of horizons must be exactly two \cite{Dymnikova}.

Of course there are other possibilities that may be considered for the exterior geometry, depending on the properties of the interior one. For example, radiating solutions may require to be matched to an exterior Vaidya \cite{vaidya} or generalized Vaidya metric \cite{vaidya-gen}.

\subsection{Regular black holes in non-linear electrodynamics}

An interesting class of regular black hole in the form \eqref{metric-rbh} can be obtained from GR coupled to a theory of non linear electrodynamics (NLED) \cite{peres,nled, nled1}. The action for GR coupled to NLED is given by
\be 
\mathcal{A}=\frac{1}{16\pi}\int d^4x\sqrt{|\mathrm{g}|}\left(\mathrm{R}-\mathcal{L}_{\rm NLED}(\mathrm{F})\right) \, ,
\ee 
where $|\mathrm{g}|$ is the determinant of the metric and the Lagrangian for NLED is
\be 
\mathcal{L}_{\rm NLED}(\mathrm{F})=\frac{4\lambda}{\alpha}\frac{(\alpha\mathrm{F})^{(\kappa+3)/4}}{[1-(\alpha \mathrm{F})^{\kappa/4}]^{1+\lambda/\kappa}} \, ,
\ee 
with $\alpha$ the coupling parameter. The Faraday tensor $\mathrm{F}_{\mu\nu}$ of Maxwell's electrodynamics gives
\be
\mathrm{F}=\mathrm{F}_{\mu\nu}\mathrm{F}^{\mu\nu} \; .
\ee
To consider a vacuum solution for the exterior we must take the energy momentum tensor as due only to $\mathcal{L}_{\rm NLED}$. For a spherically symmetric black hole coupled to NLED we have
\be 
T_{\mu\nu}=\frac{1}{4\pi}\left(\partial_{\mathrm{F}}\mathcal{L}_{\rm NLED}\mathrm{F}^\sigma_{\mu}\mathrm{F}_{\nu\sigma}-\frac{1}{4}g_{\mu\nu}\mathcal{L}_{\rm NLED}\right) \, .
\ee 

Then taking the NLED source as a magnetic charge $q_*$ and using Schwarzschild coordinates we obtain the line element in the form \eqref{metric-rbh} with
\be \label{M}
M(R)=\frac{M_0R^{\lambda}}{(R^\kappa+q^\kappa_*)^{\lambda/\kappa}} \, .
\ee 
It is obvious that the case $\lambda=0$ reduces to the Schwarzschild solution as does the case of vanishing NLED charge $q_*=0$. Also in order for the solution to be regular at $R=0$ we must evaluate the Kretschmann scalar $\mathcal{K}$ which is given by equation \eqref{K} from which we see that the condition of regularity at the center is $\lambda\geq 3$
\cite{nled4}.

Simple examples of regular black holes belonging to the above class may be obtained for $\lambda=3$ for different values of $\kappa$ \cite{bobir2}:
\begin{enumerate}
    \item For $\kappa=1$ we obtain the so-called Maxwellian black hole with
    \be 
    f=1-\frac{2M_0}{R}\left(1+\frac{q_*}{R}\right)^{-3}\;.
    \ee 
    \item For $\kappa=2$ we obtain the so-called Bardeen black hole \cite{bh1} with
    \be 
    f=1-\frac{2M_0}{R}\left(1+\frac{q_*^2}{R^2}\right)^{-3/2}\;.
    \ee
    \item For $\kappa=3$ we obtain the so-called Hayward black hole \cite{bh3} with
    \be 
     f=1-\frac{2M_0}{R}\left(1+\frac{q_*^3}{R^3}\right)^{-1}\;.
    \ee
\end{enumerate}

\section{Matching}\label{sec3}

We shall now discuss the matching of the collapsing interior to a given exterior geometry \cite{poisson}. 
The collapsing cloud is separated from the exterior spacetime by a boundary hypersurface, which in the following we will assume to follow a time-like trajectory. Junction conditions at the boundary describe the change of the matter field from the interior to the exterior, such as, for example, the separation between a dust interior from a vacuum exterior in the OSD model.
The junction conditions are obtained by assuming that the manifold $\mathcal{M}$ is divided into two distinct regions $\mathcal{M}^+$ and $\mathcal{M}^-$ separated by a
three-dimensional hypersurface $\Sigma$ and requiring that the metric be continuous across $\Sigma$ while discontinuities on $\Sigma$ may be interpreted as a matter distribution concentrated on $\Sigma$ \cite{matching1,matching2,matching3}.

Einstein's field equations hold in both regions and the line element in $\mathcal{M}^\pm$ can be written as
\begin{equation}
ds^2_\pm=g_{\mu\nu}^\pm dx^\mu_\pm dx^\nu_\pm \;,
\end{equation}
with $\{x^\mu\}^\pm$ being the coordinates in $\mathcal{M}^\pm$
($\mu, \nu=0,1,2,3$). The line element on the three-dimensional boundary surface
$\Sigma$ can be written as
\begin{equation}
ds_\Sigma^2=\gamma_{ab}dy^ady^b\;,
\end{equation}
where $\{y^a\}$ are the coordinates on the $\Sigma$ (with latin indices $a$, $b$ taking three values).
The hypersurface $\Sigma$ can be written in parametric form on both sides as
\begin{equation}
\Phi^\pm(x^\mu_\pm(y^a))=0\;.
\end{equation}
The first junction conditions then are given by the requirement that the induced metric $\gamma_{ab}$ must be the same on both
sides. Since the induced metric is
\begin{equation}
\gamma_{ab}^\pm=\frac{\partial x_{\pm}^\mu}{\partial
y^a}\frac{\partial x_{\pm}^\nu}{\partial
y^b}g_{\mu\nu}^\pm=e^\mu_{(a)}e^\nu_{(b)}g_{\mu\nu}^\pm\;,
\end{equation}
with $e^\mu_{(a)}=\partial x_{\pm}^\mu/\partial y^a$ the basis vectors tangent to $\Sigma$, in order for $\gamma_{ab}^\pm$ to be the same on both sides there must exist a coordinate transformation on $\Sigma$ for which
$\gamma_{ab}^\pm=\gamma_{ab}$
or 
\be 
[\gamma_{ab}]=\gamma_{ab}^+-\gamma_{ab}^-=0\;,
\ee 
where $[\mathcal{A}]=\mathcal{A}_+-\mathcal{A}_-$ defines the jump of a quantity $\mathcal{A}$ across $\Sigma$.
Then the metric is continuous
everywhere on $\mathcal{M}$, even though its first derivatives
might still be discontinuous across $\Sigma$.

The second junction conditions must be imposed on the first derivatives of the metric. If they are also continuous, then the hypersurface $\Sigma$ is truly a boundary. 
To evaluate these junction conditions one needs to evaluate the extrinsic curvature, also known as second fundamental form, $K_{ab}^\pm$ on both sides.
Given the unit vector $n^\mu$ normal to $\Sigma$ 
\begin{equation}
n_\mu=\frac{\partial \Phi /\partial x^\mu}{\sqrt{g^{\alpha\beta}\frac{\partial \Phi }{\partial x^\alpha}\frac{\partial \Phi }{\partial x^\beta}}}\;,
\end{equation}
the induced metric can be found from
\begin{equation}
e^\mu_a e^\nu_b\gamma^{ab}=g^{\mu\nu}-\epsilon n^\mu n^\nu\;,
\end{equation}
with $\epsilon=0$ for a null surface, $\epsilon=1$ for a spacelike surface and $\epsilon=-1$ for a timelike surface. Then the extrinsic curvature is defined as
\begin{equation}
K_{ab}=g_{\mu\nu}n^\mu\nabla_a e^\nu_{b}\;,
\end{equation}
or, expressed in coordinates,
\begin{equation}\label{SFF}
K_{ab}^\pm=\frac{\partial x_{\pm}^\mu}{\partial y^a}\frac{\partial
x_{\pm}^\nu}{\partial y^b}\nabla_\mu
n_\nu=-n_\sigma\left(\frac{\partial^2 x^\sigma}{\partial y^a
\partial y^b}+\Gamma^\sigma_{\mu\nu}\frac{\partial x^\mu}{\partial
y^a}\frac{\partial x^\nu}{\partial y^b}\right)\;.
\end{equation}

The Einstein tensor contains second derivatives of the
metric and since, as we have seen, the first derivatives may be discontinuous across $\Sigma$ this means that the second derivatives can be written as a Dirac delta on
$\Sigma$. For simplicity, let us consider a coordinate system such that $\Sigma$ is given by $x=x^3=0$. Such a coordinate always exists and the change of coordinates implies only a gauge fixing. Then the energy-momentum tensor can be written as
\begin{equation}
T_{\mu\nu}=T_{\mu\nu}^+\theta(x)+T_{\mu\nu}^-\theta(-x)+S_{\mu\nu}\delta(x)\;,
\end{equation}
where the function $\theta(x)$ is the step function
\be
\theta(x)=\left\{\begin{array}{cc}
  0, & x<0\;, \\
  1, & x>0\;,
\end{array}
\right.
\ee
for which $d\theta/dx=\delta(x)$ with $\delta$ being the Dirac delta.
Then $S_{\mu\nu}$ describes the part of the energy-momentum
tensor concentrated on $\Sigma$, which means that the components of $S_{\mu\nu}$ outside the shell $x=0$ must vanish, i.e. in the gauge used here this implies that we must have $S_{33}=S_{a3}=0$ and
\begin{equation}
S^{\mu\nu}=S^{ab}e^\mu_{(a)}e^\nu_{(b)}\;,
\end{equation}
with $a,b=0,1,2$.
From Einstein's equations we then obtain the so-called Lanczos equation as
\begin{equation}
S_{ab}=[K_{ab}]-\gamma_{ab}[K]\;,
\end{equation}
or inversely
\begin{equation}
  [K_{ab}]=S_{ab}-\frac{1}{2}\gamma_{ab}S\;.
\end{equation}

A boundary surface is defined by $S_{ab}=0$, which means that
the energy momentum tensor has a discontinuity only across the
surface. This is reflected in the extrinsic curvature for which
\begin{equation}
[K_{ab}]=0\;,
\end{equation}
implying that the first derivatives of the metric are continuous on $\Sigma$.

In the following we will consider the boundary surface for the collapsing cloud to be spherical, and assume that it will follow a time-like trajectory, although the considerations can be extended to null boundary surfaces in a rather straightforward way \cite{kijowski}.

\subsection{Spherical time-like matching}

We will now derive the first and second fundamental forms and the junction conditions for a spherical time-like shell in an arbitrary dynamical spacetime. Let's consider a generic spherical line element in the coordinates $\{x^\mu\}=\{t,r,\theta,\phi\}$ (with $\mu=0,1,2,3$) given by
\be 
d s^{2}=-A^{2} d t^{2}+B^{2} d r^{2}+C^{2} d \Omega^{2}\;.
\ee 
This is the same line element as in equation \eqref{line} with $A=e^\nu$ and $B=e^\psi$. The hypersurface $\Sigma$ of a spherical time-like boundary can be given in parametric form as
\be 
\Phi(x^\mu)=r-R_{b}(t)=0\;,
\ee 
so that the metric restricted on $\Sigma$ becomes
\be \label{Sigma-a}
d s_{\Sigma}^{2}=-\left[A^{2}-B^{2}\left(\frac{\partial R_{b}}{\partial t}\right)^{2}\right] d t^{2}+C^{2} d \Omega^{2}\;,
\ee 
where we understand that a generic function $X(t,r)$ on $\Sigma$ becomes $X_b(t)=X\left(t, R_{b}(t)\right)$, and we will omit the subscript `$b$' to avoid making the notation too cumbersome.

The metric on $\Sigma$ in coordinates $\{y^a\}=\{\tau,\theta,\phi\}$ (with $a=0,2,3$) is also
\be \label{Sigma-b}
d s_{\Sigma}^{2}=-d \tau^2+C_{b}(\tau)^{2} d \Omega^{2}\;.
\ee
Since the two line elements \eqref{Sigma-a} and \eqref{Sigma-b} must be the same we get
\bea \label{tau}
\left(\frac{d \tau}{d t}\right)^{2}&=&A^{2}-B^{2}\left(\frac{\partial R_{b}}{\partial t}\right)^{2}\;, \\
C_b(\tau)&=&C\left(t(\tau), R_{b}(t(\tau))\right)\;.
\eea 
Since we shall use the proper time on the shell $\tau$ as the trajectory's affine parameter, it is useful to invert \eqref{tau} to get
\be \label{tau2}
\left(\frac{d t}{d \tau}\right)^{2}=\frac{1}{A^{2}}\left(1+B^{2} \dot{R}_{b}^{2}\right)\;,
\ee 
with $\dot{R}_{b}=dR_b/d\tau$. Notice that in this section we use `dot' to denote derivatives with respect to the co-moving time on $\Sigma$, i.e. $\tau$, while in section \ref{sec1} we used the same notation to denote derivatives with respect the time coordinate $t$. The two will be shown to be the same for homogeneous collapse models, but one should keep in mind that they need not be the same in general.

To have a physically viable matching we need to consider also the continuity of the extrinsic curvature $K_{ab}$ on the surface which is defined by equation \eqref{SFF}.
In the case of a spherical time-like shell we get
\bea 
n_{t}&=&-\dot{R}_{b} A B\;, \\
n_{r}&=&A B \frac{\partial t}{\partial \tau}\;, \\
n_{\theta}&=&n_{\varphi}=0\;.
\eea 
A somewhat tedious calculation for the extrinsic curvature then gives
\bea 
 K_{\tau \tau}&=&-\frac{B \ddot{R}_{b}+B,_{r} \dot{R}_{b}^{2}}{\sqrt{1+B^{2} \dot{R}_{b}^{2}}}-2 \frac{\dot{R}_{b} B,_{t}}{A}-\frac{A,_{r}}{A B} \sqrt{1+B^{2} \dot{R}_{b}^{2}} \;,  \\
K_{\theta\theta}&=& C\left(\frac{B}{A}\dot{R}_bC,_t+\frac{\sqrt{1+B^2\dot{R}_b^2}}{B}C,_r\right) \;,  \\ 
K_{\phi\phi}&=&K_{\theta\theta}\sin^2\theta \;,
\eea 
where one should remember that for the purpose of the matching $A$, $B$ and $C$ must be evaluated on $\Sigma$, i.e. they are $A_b$, $B_b$ and $C_b$.

The above formalism can be used to evaluate the first and second fundamental forms for both interior and exterior by making suitable choices for $A$, $B$ and $C$.
The continuous matching between an interior a given exterior is then obtained by imposing the following junction conditions for the first and second fundamental forms
\bea \label{FFF-1}
\gamma_{\tau\tau}^+&=&\gamma_{\tau\tau}^- \;, \\ \label{FFF-2}
\gamma_{\theta\theta}^+&=&\gamma_{\theta\theta}^- \;, \\ \label{SFF-1}
K^+_{\tau\tau}&=&K^-_{\tau\tau} \;, \\ \label{SFF-2}
K^+_{\theta\theta}&=&K^-_{\theta\theta} \;.
\eea 
Of course, the conditions for $\gamma_{\phi\phi}$ and $K_{\phi\phi}$ are immediately obtained from those for $\gamma_{\theta\theta}$ and $K_{\theta\theta}$ due to spherical symmetry since we have $\gamma_{\phi\phi}= \gamma_{\theta\theta}\sin^2\theta$ and $K_{\phi\phi}= K_{\theta\theta}\sin^2\theta$.

\subsection{Interior geometry: collapse}

We shall now specialise the above treatment to some special cases. First we consider the interior $\mathcal{M}^-$ with line element \eqref{line} in coordinates $\{x^\mu\}^-=\{t,r,\theta,\phi\}$. The boundary surface $\Sigma$ given by $\Phi^-=r-r_b(t)=0$. The metric on $\Sigma$ is given by \eqref{Sigma-a} with $A^2=e^{2\nu}$ and $B^2=e^{2\psi}$.
Then for the continuity of the metric we get
\bea
\frac{d t}{d \tau}&=& e^{-\nu}\sqrt{1+e^{2\psi} \dot{r}_b^2} \;, \\
C_b(\tau)&=&C(t,r_b(t)) \;,
\eea 
with $t=t(\tau)$. Similarly for the extrinsic curvature we get the normal unit vector as
\bea 
n_{t}&=&-\dot{r}_be^{\nu+\psi} \;, \\
n_{r}&=&e^{2\psi}\sqrt{e^{-2\psi}+\dot{r}_b^2} \;, \\
n_\theta&=&n_\phi=0 \;,
\eea 
so that
\bea
K_{\tau \tau}^{-} &=& -\frac{\ddot{r}_b+\psi'\dot{r}_b^2}{\sqrt{e^{-2\psi}+\dot{r}_b^2}}-2\dot{r}_b\psi,_te^{\psi-\nu}-\nu'\sqrt{e^{-2\psi}+\dot{r}^2_b} \;, \\
K_{\vartheta \theta}^{-} &=& C\left(C,_t\dot{r}_be^{\psi-\nu}+C'\sqrt{e^{-2\psi}+\dot{r}_b^2}\right) \;,
\eea 
where we used primed quantities for derivatives with respect to $r$ but kept the subscript $X,_t$ for derivatives with respect to $t$ and dotted quantities for derivatives with respect to $\tau$.
In the case of a co-moving boundary $r_b=\text{const}.$ the above equations reduce to
\bea \label{fff1}
\frac{d t}{d \tau}&=&e^{-\nu} \;, \\ \label{fff2}
C(t,r_b)&=&C_b(\tau) \;, \\ \label{sff1}
K_{\tau \tau}^{-} &=& \nu'e^{-\psi} \;, \\ \label{sff2}
K_{\vartheta \theta}^{-} &=& CC'e^{-\psi} \;.
\eea 
As we shall see later, for homogeneous dust collapse we may take $\nu=0$ and therefore identify $t$ with $\tau$.

\subsection{Exterior geometry: regular black holes}

If we consider the exterior geometry $\mathcal{M}^+$ with line element \eqref{metric-rbh} in Schwarzschild coordinates $\{x^\mu\}^+=\{T,R,\theta,\phi\}$, thus setting $A^2=f$ and $B^2=1/f$, the boundary surface $\Sigma$ can be given by $\Phi_+=R-R_{b}(T)=0$, with $R_b(T)$ being the trajectory of a radial infalling particle in the spacetime. Then for the continuity of the metric we get
\bea 
\frac{d T}{d \tau}&=&\frac{\sqrt{f+\dot{R}_{b}^{2}}}{f} \;, \\ 
C_b(\tau)&=&R_b(T(\tau)) \;,
\eea 
with $T=T(\tau)$.
For the extrinsic curvature we have
\bea
K_{\tau \tau}^{+} &=&-\frac{1}{\sqrt{f+\dot{R}_{b}^{2}}}\left(\ddot{R}_{b}+\frac{f_{, R}}{2}\right) \;, \\
K_{\vartheta \theta}^{+} &=&R_{b} \sqrt{f+\dot{R}_{b}^{2}} \;.
\eea 

In the case of a regular black hole with $M(R)$ given by equation \eqref{M} we then have
\be 
f,_R=\frac{2M(R)}{R^2}-\frac{2M,_R}{R}=\frac{2M_0R^{\lambda-2}}{(R^\kappa+q^\kappa_*)^{\lambda/\kappa}}\frac{1+(1-\lambda)q_*^\kappa/R^\kappa}{1+q_*^\kappa/R^\kappa}\;.
\ee 

The Schwarzschild case is immediately obtained from the regular black hole case if we impose $M=M_0=\text{const}.$ or $\lambda=0$.
We then get
\bea \label{match-schw1}
\frac{d T}{d \tau}&=&\frac{\sqrt{1-2M_0/R_b+\dot{R}_{b}^{2}}}{1-2M_0/R_b} \;, \\ \label{match-schw2}
C_b(\tau)&=&R_b(T(\tau)) \;, \\ \label{match-schw3}
K_{\tau \tau}^{+} &=&-\frac{\ddot{R}_{b}+M_0/R_b^2}{\sqrt{1-2M_0/R_b+\dot{R}_{b}^{2}}} \;, \\ \label{match-schw4}
K_{\vartheta \theta}^{+} &=&R_{b} \sqrt{1-2M_0/R_b+\dot{R}_{b}^{2}} \;.
\eea 
In the following we will consider semi-classical homogeneous dust collapse models for the interior and investigate under what conditions they may be matched to exterior solutions given by regular black holes in GR coupled to NLED.

\section{Dust collapse}\label{sec4}

Dust collapse can be obtained from equations \eqref{EE-rho2}-\eqref{misner-3} by setting $p=0$. Then equation \eqref{EE-p2} becomes $\dot{m}=0$ and implies that $m=m(r)$ while equation \eqref{nu2} gives $\nu=\nu(t)$ which can be set to $\nu=0$ by a suitable rescaling of the co-moving time
\cite{chapter}.  
Equation \eqref{EE-G} becomes $\dot{G}=0$ from which we get
$b=b(r)$ in equation \eqref{EE-G2}. Then from equation \eqref{misner-3} we finally get
\be \label{eom2}
\dot{a}^2=\frac{m}{a}+b\; ,
\ee 
from which we see that in the case of inhomogeneous dust we must have $a=a(t,r)$. 
The Kretschmann scalar for dust collapse is
\be
\mathcal{K}=48\frac{m^2}{a^6}-32\frac{m(3m+rm')}{a^5(a+ra')}+12\frac{(3m+rm')^2}{a^4(a+ra')^2} \;,
\ee 
which diverges for $a=0$ signalling the occurrence of the singularity. The complete set of Einstein's equations for inhomogeneous dust collapse may be complicated to solve analytically \cite{joshi2,joshi,JD} and can require the aid of numerical methods\cite{mw,musco}. For example, in inhomogeneous models one has to consider the relative trajectories of different shells, which may overlap leading to `shell crossing' singularities \cite{cross1,cross2,cross3,cross4}. Therefore applying semi-classical corrections to inhomogeneous dust may be complicated \cite{liu,bojo} and one may look at the homogeneous case as a more manageable toy model.

\subsection{Homogeneous dust}

The OSD model describes homogeneous dust collapse and is obtained by further requiring that $\epsilon=\epsilon(t)$, i.e. the density is homogeneous. Imposing homogeneity in equation \eqref{EE-rho2} implies that $m'=0$ and $a'=0$ and therefore $m=m_0={\rm const.}$ and $a=a(t)$. The energy density is then simply given by
\be 
\epsilon (t)=\frac{3m_0}{a^3}\; .
\ee 
From equation \eqref{eom2} with $a=a(t)$ and $m=m_0$ we then get the additional condition $b=k={\rm const.}$. 
We can then restrict the allowed values of $k$ to $k=0,\pm 1$ via the additional rescaling $r\rightarrow r/\sqrt{k}$, $t\rightarrow t/\sqrt{k}$. The case $k=0$ is called `marginally bound' and it corresponds to infalling particles having zero velocity at spatial infinity. The case $k=1$ is called `unbound' and it corresponds to infalling particles having positive velocity at spatial infinity. Finally the case $k=-1$ is called `bound' collapse and it corresponds to infalling particles reaching zero velocity at a finite radius.
The system is fully solved once we find the solution of the equation \eqref{eom2}, written in the form
\be \label{eom}
\dot{a}=-\sqrt{\frac{m_0}{a}+k}\; ,
\ee 
with the minus sign chosen in order to describe collapse.

In the marginally bound case, given by $k=0$, with initial condition $a(0)=1$, the above equation is immediately integrated to give
\be \label{a(t)}
a(t)=\left(1-\frac{3}{2}\sqrt{m_0}t\right)^{2/3} \; ,
\ee 
and the singularity is reached for $a=0$ in a finite comoving time $t_s=2/(3\sqrt{m_0})$.
In the bound and unbound cases we obtain the solution in parametric form as
\be 
a(t)= \frac{m_0}{2}(1-\cos\eta)\; ,
\ee
with
\be
\eta-\sin\eta=\frac{2}{m_0}(t-t_s)\; ,
\ee
if $k=-1$ and
\be 
a(t)= \frac{m_0}{2}(\cosh\eta-1)\; ,
\ee 
with
\be 
\sinh\eta-\eta=\frac{2}{m_0}(t-t_s)\; ,
\ee
if $k=+1$.
The line element then takes the simple form
\be \label{metric-HD}
ds^2=-dt^2+a^2\left(\frac{dr^2}{1+kr^2}+r^2d\Omega^2\right)\; ,
\ee 
and the Krestchmann scalar reduces to
\be
\mathcal{K}=60\frac{m_0^2}{a^6} \; .
\ee 

The matching must be done with the Schwarzschild metric in the exterior since there is no inflow or outflow of matter through any shell $r$ of the interior all the way up to the boundary $r_b={\rm const}.$.
In order to perform the matching with the interior for any value of $k=0,\pm 1$ we need to rewrite the line element with the following change of coordinates
\be 
r=\left\{\begin{matrix}
\sin\zeta \; \text{  for  } \; k=-1\; ,\\ 
\zeta \; \text{  for  } \; k=0\; ,\\ 
\sinh\zeta \; \text{  for  } \; k=+1 \; ,
\end{matrix}\right.
\ee 
which gives the line element \eqref{metric-HD} as
\be 
ds^2=-dt^2+a^2\left(d\zeta^2+r(\zeta)^2d\Omega^2\right)\;,
\ee 
and consider $\zeta$ as the radial coordinate with the boundary at $r_b=\sin\zeta_b$ for $k=1$ and $r_b=\sinh\zeta_b$ for $k=-1$.

In the case of homogeneous dust collapse the first and second fundamental forms are immediately obtained from equations \eqref{fff1}-\eqref{sff2} and give
\bea
\frac{d t}{d \tau}&=&1 \;, \\
C(t,r_b)&=&r_ba(t) \;, \\
K_{\tau \tau}^{-} &=& 0 \;, \\
K_{\vartheta \theta}^{-} &=& r_ba(t) \;.
\eea 
The first and second fundamental forms for the Schwarzschild case are given by equations \eqref{match-schw1}-\eqref{match-schw4} so that continuity of the metric gives the Schwarzschild time $T$ as a function of the co-moving time $t$ from
\be
\frac{d T}{d t}=\frac{\sqrt{1-2M_0/R_b+\dot{R}_{b}^{2}}}{1-2M_0/R_b} \;,
\ee
and the trajectory of the collapsing boundary as
\be \label{match}
R_b(T(t))=r_ba(t) \;.
\ee
The matching of the extrinsic curvature then gives the two additional relations
\bea \label{Rddot}
0 &=&-\frac{\ddot{R}_{b}+M_0/R_b^2}{\sqrt{1-2M_0/R_b+\dot{R}_{b}^{2}}} \;, \\
r_ba(t) &=&R_{b} \sqrt{1-2M_0/R_b+\dot{R}_{b}^{2}} \;,
\eea 
which can be rewritten as
\bea 
\ddot{R}_b&=&-\frac{M_0}{R_b^2} \;, \\
1&=&\sqrt{1-2M_0/R_b+\dot{R}_{b}^{2}} \;.
\eea 
It is easy to see that these two equations are equivalent, since by squaring the second equation and deriving with respect to $t$ we obtain the first one. Also equation \eqref{Rddot} is equivalent to the equation of motion \eqref{eom} evaluated at the boundary if we make use of the matching condition \eqref{match} and impose that
\be 
2M_0=F(r_b)=m_0r_b^3 \; .
\ee 
The above discussion shows that for homogeneous dust the trajectory of a particle at the boundary of the cloud is a geodesic determined by the amount of matter contained within it. We should then be able to write the same equation of motion for a particle on the boundary using the interior or the exterior metric. This can be done by using Lema\`itre coordinates for the exterior. 

\subsection{Schwarzschild in Lema\`itre coordinates}

In the case of dust collapse the absence of pressures implies that each particle at a co-moving radius $r_0$ must follow the geodesic determined by the matter content present within $r\leq r_0$. This is the same geodesic followed by a particle in radial free fall in the Schwarzschild geometry with the mass parameter given by the amount of mass contained within $r\leq r_0$. 
One easy way to illustrate the above idea is to consider for the exterior a set of coordinates used by an observer in free fall. These coordinates, known as Lema\`itre coordinates $\{\tau,\rho\}$ \cite{L}, can be defined for a general exterior of the form \eqref{metric-rbh}, which includes Schwarzschild.
Lema\`itre coordinates are obtained from the two transformations $R=R(\tau,\rho)$, $T=T(\tau,\rho)$ given by
\bea 
d\tau&=& dT+\frac{g(R)}{f(R)}dR \, , \\
d\rho&=&dT+\frac{1}{g(R)f(R)}dR \, ,
\eea 
with $g=\sqrt{1-f}$ and $f$ given by \eqref{F-nled},
so that for Schwarzschild we get the line element as
\be \label{metric1}
ds^2=-d\tau^2+\frac{2M_0}{R}d\rho^2+R(\tau,\rho)^2d\Omega^2 \, .
\ee

A particle in radial free fall in the Schwarzschild geometry in Lema\`itre coordinates is then located at $\rho=\rho_0$, $\theta=\theta_0$ and $\phi=\phi_0$. Then we can find the particle's trajectory as $R_0(\tau)=R(\tau,\rho_0)$ from the change of coordinates, since from
\be 
d\rho-d\tau=\frac{1}{g}dR=\sqrt{\frac{R}{2M_0}}dR \, ,
\ee 
evaluated at $\rho=\rho_0$, i.e. $d\rho_0=0$, we get
\be \label{0}
\frac{dR_0}{d\tau}=-\sqrt{\frac{2M_0}{R_0}} \, .
\ee
Integrating the above equation with the initial condition $R_0(0)=\rho_0$ gives the trajectory as
\be \label{lem}
R_0(\tau)=\rho_0\left(1-\frac{3}{2}\sqrt{\frac{2M_0}{\rho_0^3}}\tau\right)^{2/3}=\rho_0a(\tau) \, ,
\ee 
where the adimensional function $a(\tau)$ is the scale factor and we can define $2M_0=m_0\rho_0^3$ which then relates to the matching with the interior.
Notice that, as expected, equation \eqref{0} is identical to the equation of motion for marginally bound homogeneous dust collapse. Notice that the cases of bound and unbound collapse may also be obtained in a similar manner by considering a particle in radial free fall with an energy per unit mass such at its initial velocity zero at a finite radius $R$ or is positive at spatial infinity. In fact if we call $E$ the energy per unit mass of the test particle its velocity is
\be \label{00}
\frac{dR_0}{d\tau}=-\sqrt{\frac{2M_0}{R_0}+E^2-1} \, .
\ee 
which gives the bound (unbound) case for $k=E^2-1<0$ ($E^2>1$, respectively). 

\subsection{Semi-classical homogeneous dust}

Semi-classical corrections to the energy momentum tensor can be introduced to model repulsive effects at large curvature which may lead to the singularity resolution. In general one may consider an effective density given in the form \eqref{eps-eff}. However, the choice of $\epsilon_{\rm eff}$ must be well motivated by some approach to gravity at large curvatures. Also, in order to avoid an unnecessary proliferation of arbitrary parameters, it may be wise to impose that $\alpha_i$ depend upon one single free parameter for all $i$.
One simple model for a modification of the OSD scenario, inspired by Loop Quantum Gravity (LQG) was proposed in \cite{collapse6}. The effective energy density in this model is given by 
\be \label{LQG-epsilon}
\epsilon_{\rm eff}=\epsilon\left(1-\frac{\epsilon}{\epsilon_{\rm cr}}\right)\; ,
\ee
where $\epsilon_{\rm cr}=3m_0/a_{\rm cr}^3$ is a critical density scale related to $\alpha_1$ in equation \eqref{eps-eff} via $\alpha_1=-1/\epsilon_{\rm cr}$. Notice that $\alpha_i=0$ for $i>1$. The corresponding equation of motion obtained from equation \eqref{misner-3} is
\be \label{LQG-adot}
\dot{a}=-\sqrt{\frac{m_0}{a}\left(1-\frac{a_{\rm cr}^3}{a^3}\right)+k}\; .
\ee 

The above model can be easily generalised if we consider
\be 
\epsilon_{\rm eff}=\epsilon\left[1\pm\left(\frac{\epsilon}{\epsilon_{\rm cr}}\right)^\beta\right]^\gamma \, ,
\ee
for which the equation of motion becomes
\be \label{semi-eom}
\dot{a}=-\sqrt{\frac{m_0}{a}\left(1\pm\frac{a_{\rm cr}^{3\beta}}{a^{3\beta}}\right)^\gamma+k} \, .
\ee
Since we wish to retain terms in $\epsilon^2$ in the expansion of the effective energy momentum, namely keeping $\alpha_1\neq 0$, we must choose $\beta=1$.

Different scenarios are possible depending on the sign chosen and the values of $\gamma$ and $k$. In brief, a singularity will develop if $a\rightarrow 0$ while $\dot{a}$ remains negative. On the other hand, collapse will halt if $\dot{a}\rightarrow 0$ while $a$ remains finite. In this case collapse will turn into expansion if $\ddot{a}\neq 0$ when $\dot{a}=0$. Therefore it is clear that in order to understand the behavior of such models one has to solve equation \eqref{semi-eom} and study $a$, $\dot{a}$ and $\ddot{a}$.

The question now is whether this kind of models can be obtained from some approach to modify GR at large curvatures.
If we describe the semi-classical corrections to dust collapse in the form of an effective energy momentum tensor, then such energy momentum must carry to the exterior and affect the geometry outside the collapsing sphere. We must then 
investigate the conditions under which the collapsing interior described by the equation of motion \eqref{semi-eom} can be matched to a suitable exterior. In the following we will take a regular black hole exterior obtained from GR coupled to NLED of the form \eqref{metric-rbh} with $M(R)$ in the form given in equation \eqref{M}.

To perform the matching of the above models we need to apply the junction conditions developed in section \ref{sec3}. Homogeneity of collapse implies again that we can identify the co-moving time $t$ with the proper time on $\Sigma$. Then the relation between $T$ in the exterior and $t$ is given by
\be 
\frac{d T}{d t}=\frac{\sqrt{f+\dot{R}_{b}^{2}}}{f} \;,
\ee 
and junction conditions for the metric in the case of a constant co-moving boundary $r=r_b$ imply simply
\be  
r_ba(t)=R_b(T(t)) \;.
\ee
The remaining junction conditions for the extrinsic curvature become
\bea
0 &=&-\frac{1}{\sqrt{f+\dot{R}_{b}^{2}}}\left(\ddot{R}_{b}+\frac{f_{, R}}{2}\right) \;, \\
r_ba(t) &=&R_{b} \sqrt{f+\dot{R}_{b}^{2}} \;.
\eea 
which again reduce to two equivalent equations, namely
\bea
\ddot{R}_{b}&=&-\frac{f_{, R}}{2} \;, \\ \label{nled-junct}
1 &=&f+\dot{R}_{b}^{2} \;.
\eea
as it can easily be seen by differentiating the second one with respect to $t$.
Equation \eqref{nled-junct} with $f$ given by equation \eqref{metric-rbh} is formally identical to equation \eqref{semi-eom} and so we see that the junction conditions are fully satisfied once we identify
\bea 
2M_0&=&r_b^3m_0 \;, \\
q_*^\kappa&=&(r_ba_{\rm cr})^{3\beta} \;, \\
\frac{\lambda}{\kappa}&=&-\gamma \; .
\eea  

\subsection{NLED black holes in Lemaitre coordinates}

Similarly to what we did for homogeneous dust and Schwarzschild in Lema\`itre coordinates we can follow the same steps for semi-classical homogeneous dust and regular black holes. To move to Lema\`itre coordinates $\{\tau,\rho\}$ we define
\be 
g(R)=\sqrt{\frac{2M(R)}{R}}=\sqrt{\frac{2M_0R^{\lambda-1}}{(R^\kappa+q_*^\kappa)^{\lambda/\kappa}}} \, .
\ee
The line element can then be written in the form 
\be 
ds^2=-d\tau^2+\frac{2M(R)}{R}d\rho^2+R(\tau,\rho)^2d\Omega^2 \, ,
\ee
and the change of coordinates gives
\be 
d\rho-d\tau=\frac{1}{g}dR=\sqrt{\frac{R}{2M(R)}}dR \, .
\ee 
A free falling observer at $\rho=\rho_0=\text{const}.$ with constant values for $\theta$ and $\phi$ follows the trajectory $R_0(\tau)=R(\tau,\rho_0)$ and must satisfy the equation of motion
\be \label{1}
\frac{dR_0}{d\tau}=-\sqrt{\frac{2M_0}{R_0}\left(1+\frac{q_*^\kappa}{R_0^\kappa}\right)^{-\lambda/\kappa}} \, .
\ee 
With the scaling $R_0(\tau)=\rho_0a(\tau)$, $2M_0=m_0\rho_0^3$ and defining $q_*=\rho_0q$ the above equation becomes
\be \label{2}
\frac{da}{d\tau}=-\sqrt{\frac{m_0}{a}\left(1+\frac{q^\kappa}{a^\kappa}\right)^{-\lambda/\kappa}} \, ,
\ee 
which resembles the equation of motion of semi-classical dust collapse \eqref{semi-eom} for $k=0$. In fact the two equations coincide if we make use of the junction conditions and identify $\gamma=-\lambda/\kappa$ and $\kappa=3\beta$. Also notice that in the collapse model $a_{\rm cr}>0$ while in principle $q$ may be positive or negative. Having fixed $\beta=1$ implies that we can take $q>0$ and consider the case of negative $q$ by changing the sign in front of $q^\kappa/a^\kappa$, thus retrieving exactly equation \eqref{semi-eom} in the marginally bound case. The bound and unbound cases are again obtained by writing equation \eqref{1} with a different energy per unit mass for the test particle, which corresponds to a different velocity, similarly to equation \eqref{00}.

We thus have established a correspondence between a class of regular black holes in NLED and a class of semi-classical homogeneous dust collapse models. However, we know that not all values of $\lambda$ give a regular black hole solution and thus we need now to investigate in detail the possible endstates of collapse depending on the sign of the NLED charge q and the parameter $\lambda$. To keep the analysis as general as possible we will also allow for collapse to be bound, unbound or marginally bound, thus reintroducing the parameter $k$ in the equation of motion.

\subsection{Final fates}

To study the qualitative behavior of the class of semi-classical dust collapse models obtained from GR coupled to NLED let's consider the one dimensional dynamical system given by equation \eqref{semi-eom} with $\beta=1$, namely
\be \label{din}
\dot{a}=j(a)=-\sqrt{\frac{1}{a}\left(1\pm\frac{q^3}{a^3}\right)^\gamma+k} \, ,
\ee 
with $a>0$, and where we have normalized the scale factor by substituting $a\rightarrow m_0a$ and $q\rightarrow m_0q$. Therefore the assumption that $q<<m_0$ now becomes $q<<1$. The rescaling also implies that for negative $k$ we must take $k>-1$ in order for $j$ to be real at initial time. 
Since $j\leq 0$ the system moves towards smaller values of $a$ from the initial state which may be taken as $a(t_i)=1$, for which $j(1)=-\sqrt{(1\pm q^3)^\gamma+k}$. Since the system is one-dimensional the only possible outcome is that $a$ goes to a fixed point in a finite or infinite amount of time. In the marginally bound case, the fixed point being zero in the case of a plus sign and $q$ in the case of minus sign. For simplicity, in the following we shall denote as $(+)$ the case with plus sign and $(-)$ the case with minus sign in equation \eqref{din}.
Accordingly, the only possible outcomes are that $j$ goes to $-\infty$, zero or a finite, non zero, value and such outcomes may be reached as $a$ goes to zero or as $a$ goes to a finite value.

For each value of $k$ we can distinguish scenarios based on the sign in equation \eqref{din} and the value of $\gamma$. Notice that for $k<0$ we will have a zero of $j(a)$ at large $a$, which could be taken as an initial condition, and the dynamics will occur for smaller $a$ where $j$ could potentially have another zero. The results are summarised as follows
\begin{enumerate}
    \item  Sign $(+)$ and $k>0$: 
    \begin{itemize}
        \item $\gamma<-1/3 \Rightarrow j \xrightarrow[a\to 0]{} -\sqrt{k}$.
        \item $\gamma=-1/3 \Rightarrow j \xrightarrow[a\to 0]{} -\sqrt{k+1/q}$.
        \item $\gamma>-1/3 \Rightarrow j \xrightarrow[a\to 0]{} -\infty$.
    \end{itemize}
    \item  Sign $(-)$ and $k>0$:
    \begin{itemize}
        \item $\gamma< 0\Rightarrow j \xrightarrow[a\to q]{} -\infty$.
        \item $\gamma=0 \Rightarrow j \xrightarrow[a\to 0]{} -\infty$.
        \item $\gamma>0$ not even $\Rightarrow j \xrightarrow[a\to a^*]{} 0$ with $a^*<q$.
        \item $\gamma>0$ and even $\Rightarrow j \xrightarrow[a\to 0]{} -\infty$.
    \end{itemize}
    \item  Sign $(+)$ and $k=0$:
    \begin{itemize}
        \item $\gamma<-1/3 \Rightarrow j \xrightarrow[a\to 0]{} 0$.
        \item $\gamma=-1/3 \Rightarrow j \xrightarrow[a\to 0]{} -1/\sqrt{q}$.
        \item $\gamma>-1/3 \Rightarrow j \xrightarrow[a\to 0]{} -\infty$.
    \end{itemize}
    \item  Sign $(-)$ and $k=0$:
    \begin{itemize}
        \item $\gamma< 0 \Rightarrow j \xrightarrow[a\to q]{} -\infty$.
        \item $\gamma= 0 \Rightarrow j \xrightarrow[a\to 0]{} -\infty$.
        \item $\gamma> 0$ not even $ \Rightarrow j \xrightarrow[a\to q]{} 0$ .
        \item $\gamma>0$ and even $\Rightarrow j \xrightarrow[a\to 0]{} -\infty$\footnote{Notice that $j(q)=0$ in this case.}.
    \end{itemize}
    \item  Sign $(+)$ and $k<0$:
    \begin{itemize}
        \item $\gamma<-1/3  \Rightarrow j \xrightarrow[a\to a^*]{} 0$ with $a^*<q$.
        \item $\gamma=-1/3  \Rightarrow j \xrightarrow[a\to 0]{} -\sqrt{k+1/q}$.
        \item $\gamma>-1/3  \Rightarrow j \xrightarrow[a\to 0]{} -\infty$.
    \end{itemize}
    \item  Sign $(-)$ and $k<0$:
    \begin{itemize}
        \item $\gamma<0 \Rightarrow j \xrightarrow[a\to q]{} -\infty$.
        \item $\gamma=0 \Rightarrow j \xrightarrow[a\to 0]{} -\infty$.
        \item $\gamma>0 \Rightarrow j \xrightarrow[a\to a^*]{} 0$ with $a^*<q$.
    \end{itemize}
\end{enumerate}

The acceleration is another important element to determine the final outcome. Differentiating equation \eqref{din} with respect to $t$ we find
\be 
\ddot{a}=-\frac{1}{2}\frac{(a^3\pm q^3)^\gamma}{a^{3\gamma+2}}\left(1\pm\frac{3\gamma q^3}{a^3\pm q^3}\right)\; ,
\ee 
from which we can easily see that for $(+)$ we have 
\begin{itemize}
    \item $\ddot{a}\xrightarrow[a\to 0]{} 0$ if $\gamma<-2/3$, 
    \item $\ddot{a}\xrightarrow[a\to 0]{} q^2/2$ if $\gamma=-2/3$,
    \item $\ddot{a}\xrightarrow[a\to 0]{} -\infty$ if $\gamma>-2/3$,
\end{itemize}
while for $(-)$ we have 
\begin{itemize}
    \item $\ddot{a}\xrightarrow[a\to q]{} -\infty$ if $\gamma<1$,
    \item $\ddot{a}\xrightarrow[a\to q]{} 3/(2q^2)$ if $\gamma=1$,
    \item $\ddot{a}\xrightarrow[a\to q]{} 0$ if $\gamma>1$.
\end{itemize}

For example, the OSD case, i.e. $\gamma=0$, we have $j$ going to minus infinity as $a$ goes to zero and $\ddot{a}$ also goes to minus infinity. On the other hand for $\gamma=1$ with $(-)$ we have that $j$ goes to zero as $a$ goes to a finite value and $\ddot{a}$ goes to a finite value. Also, for $\gamma=-1$ in the $(+)$ case we have that $j$ goes to zero if $k\leq 0$ and $\ddot{a}$ goes to zero leading asymptotically to an equilibrium configuration.



\begin{figure*}[tt]
\begin{minipage}{0.49\linewidth}
\begin{overpic}
[width=0.97\linewidth]{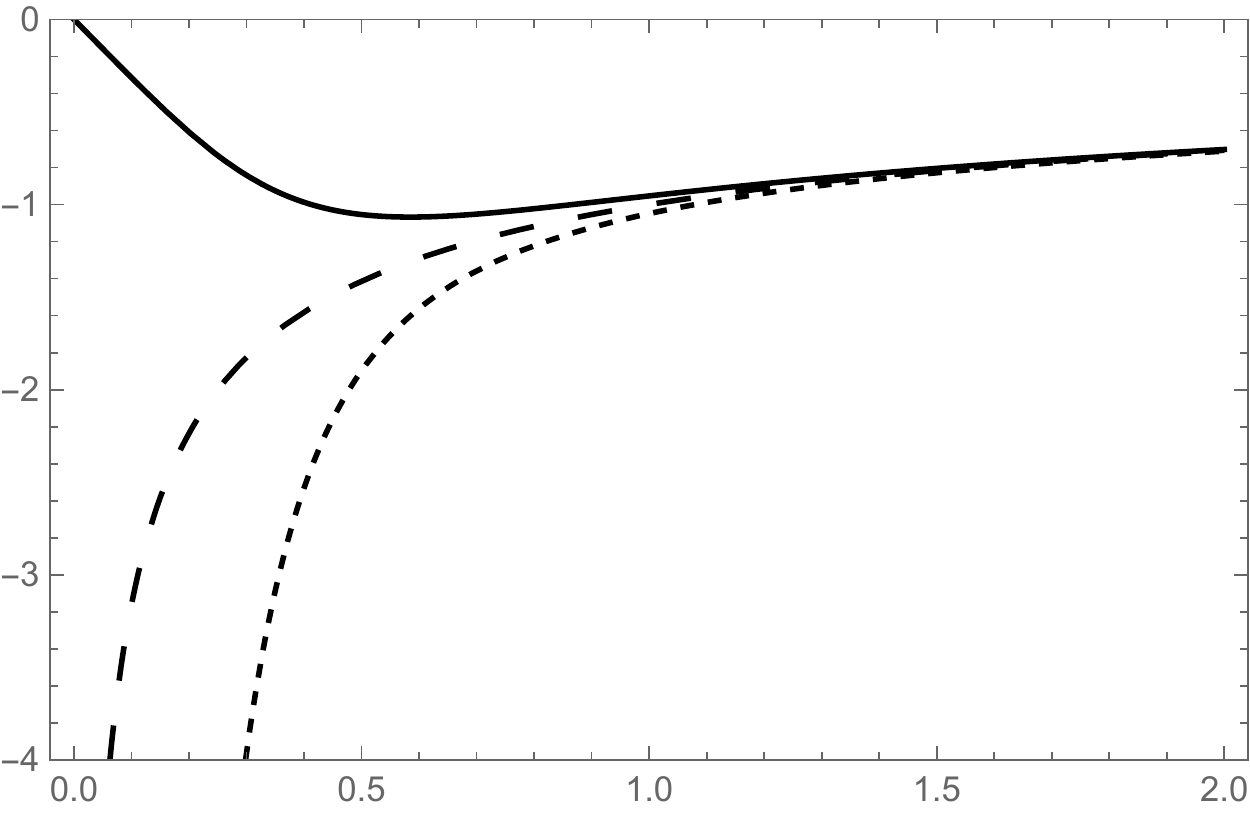}
\put(6,50){$j(a)$}
\put(90,6){$a$}
\end{overpic}
\end{minipage}
\hfill 
\begin{minipage}{0.49\linewidth} 
\begin{overpic}
[width=0.97\linewidth]{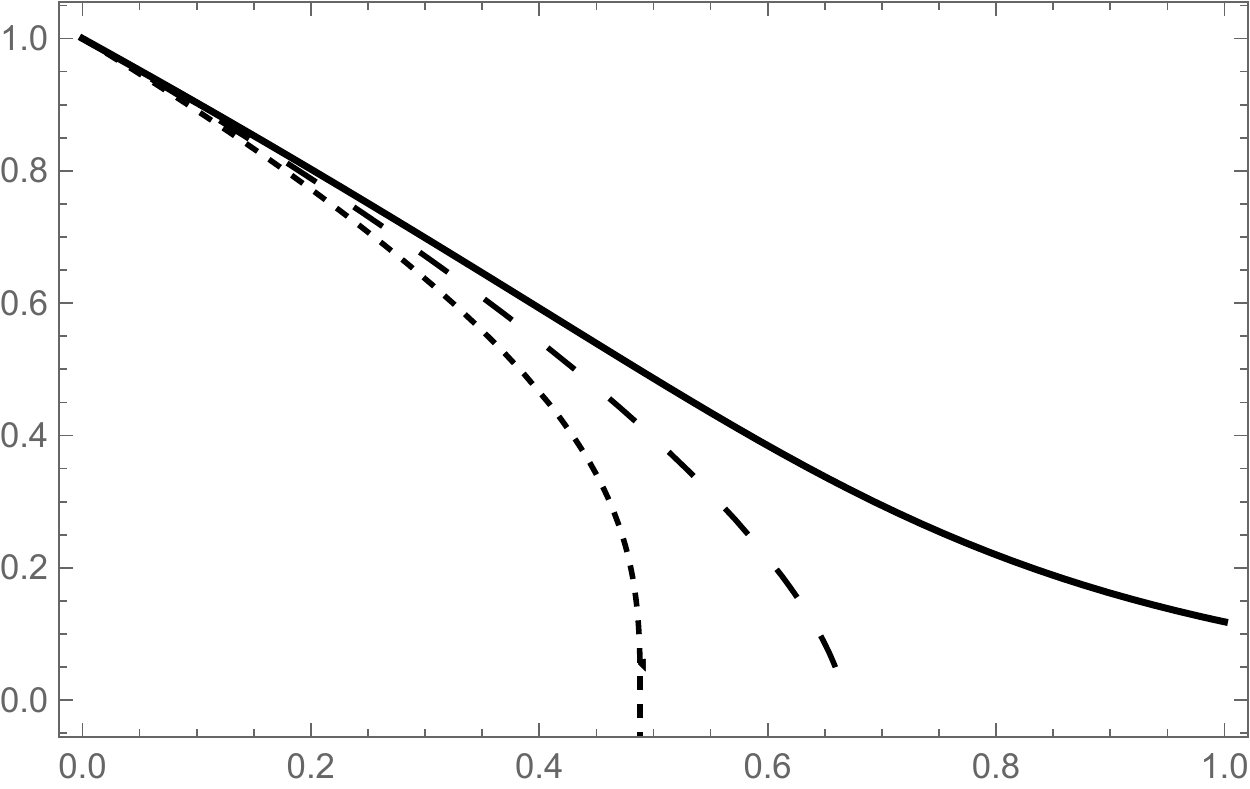}
\put(6,50){$a(t)$}
\put(90,6){$t$}
\end{overpic}
\end{minipage}
\caption{Semi-classical marginally bound collapse, i.e. $k=0$, with $(+)$ in equation \eqref{din}. Left panel: The function $j(a)=\dot{a}$ is plotted for $\gamma=0$ (large dashed), $\gamma=1$ (short dashed) and $\gamma=-1$ (solid). For the sake of clarity the range of the horizontal axis is $(0,2)$ even if the initial condition is taken as $a(t_i)=1$. Right panel: The scale factor $a(t)$ is plotted for $\gamma=0$ (large dashed), $\gamma=1$ (short dashed) and $\gamma=-1$ (solid). The case $\gamma=0$ corresponds to OSD collapse, while $\gamma=-1$ corresponds to collapse leading to the Hayward black hole. The plots are obtained for $q^3=0.1$.} \label{plus-k0}
\end{figure*}

\begin{figure*}[ht]
\begin{minipage}{0.49\linewidth}
\begin{overpic}
[width=0.97\linewidth]{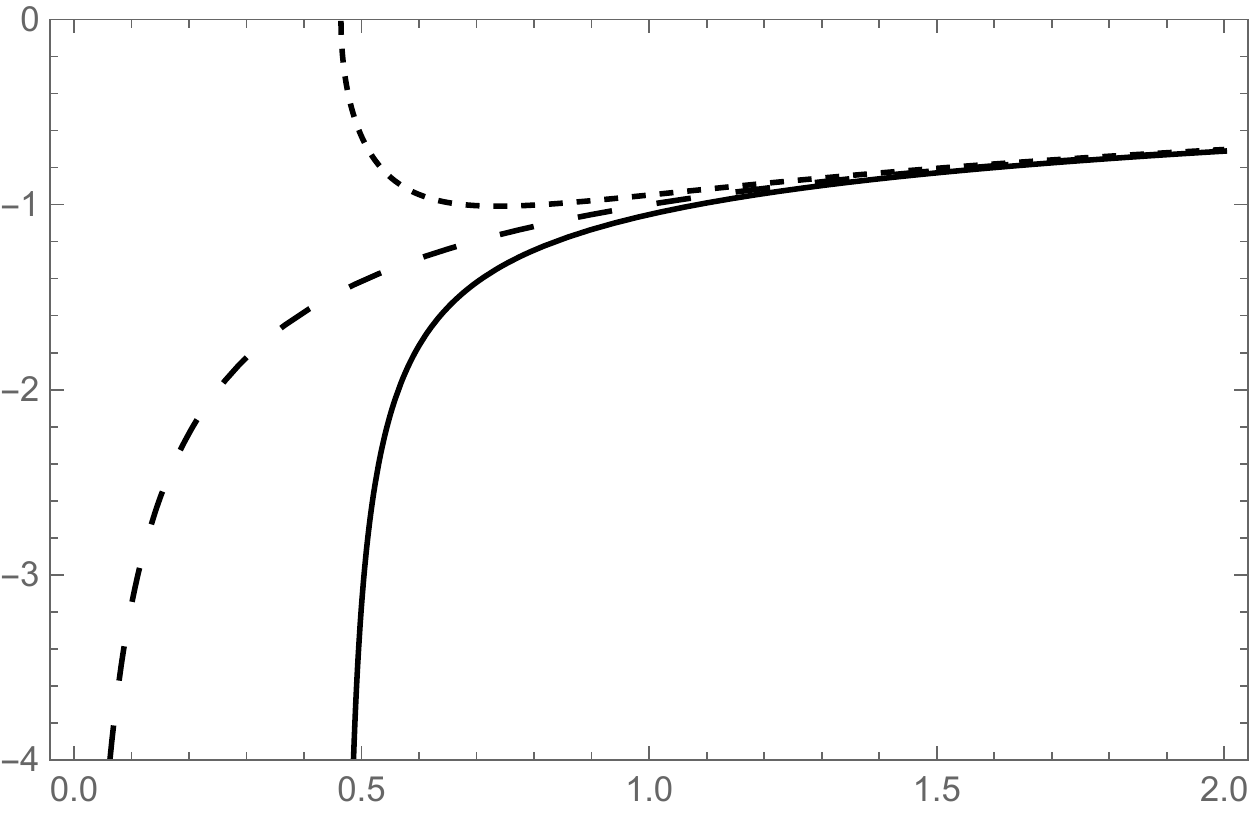}
\put(6,50){$j(a)$}
\put(90,6){$a$}
\end{overpic}
\end{minipage}
\hfill 
\begin{minipage}{0.49\linewidth}
\begin{overpic}
[width=0.97\linewidth]{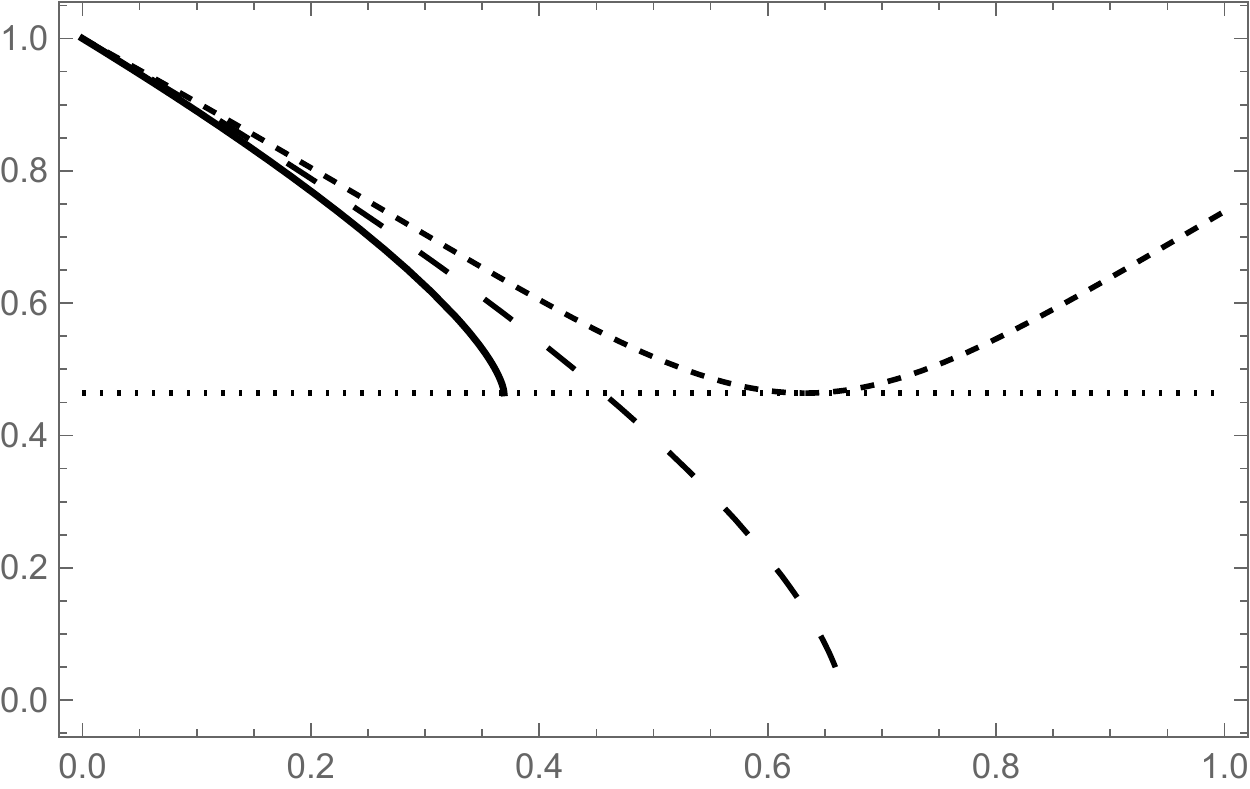}
\put(6,50){$a(t)$}
\put(90,6){$t$}
\end{overpic}
\end{minipage}
\caption{Semi-classical marginally bound collapse, i.e. $k=0$, with $(-)$ in equation \eqref{din}. Left panel: The function $j(a)=\dot{a}$ is plotted for $\gamma=0$ (large dashed), $\gamma=1$ (short dashed) and $\gamma=-1$ (solid). For the sake of clarity the range of the horizontal axis is $(0,2)$ even if the initial condition is taken as $a(t_i)=1$. Right panel: The scale factor $a(t)$ is plotted for $\gamma=0$ (large dashed), $\gamma=1$ (short dashed) and $\gamma=-1$ (solid). The case $\gamma=0$ corresponds to OSD collapse, while $\gamma=1$ corresponds to the LQG inspired bounce model described in \cite{collapse6}. The plots are obtained for $q^3=0.1$.} \label{minus-k0}
\end{figure*}

\begin{figure*}[tt]
\begin{minipage}{0.49\linewidth}
\begin{overpic}
[width=0.97\linewidth]{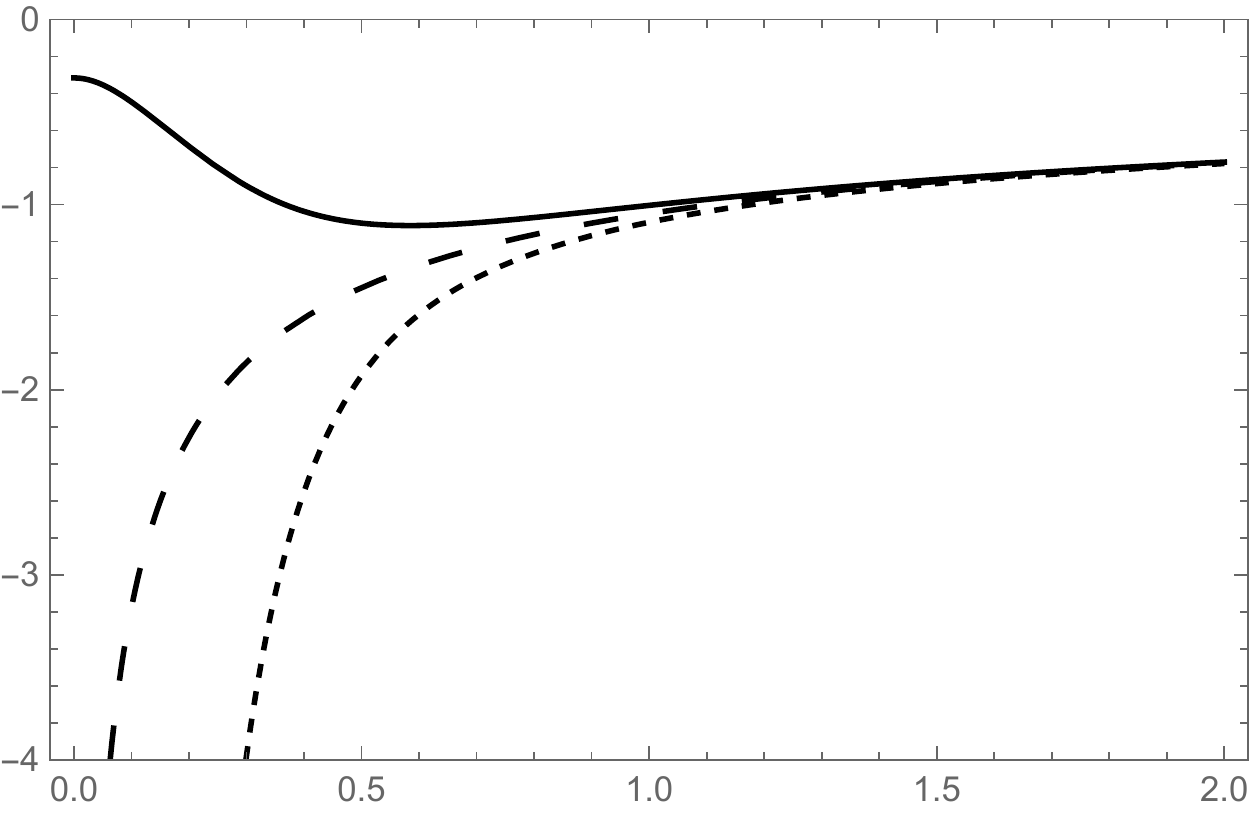}
\put(6,50){$j(a)$}
\put(90,6){$a$}
\end{overpic}
\end{minipage}
\hfill 
\begin{minipage}{0.49\linewidth}
\begin{overpic}
[width=0.97\linewidth]{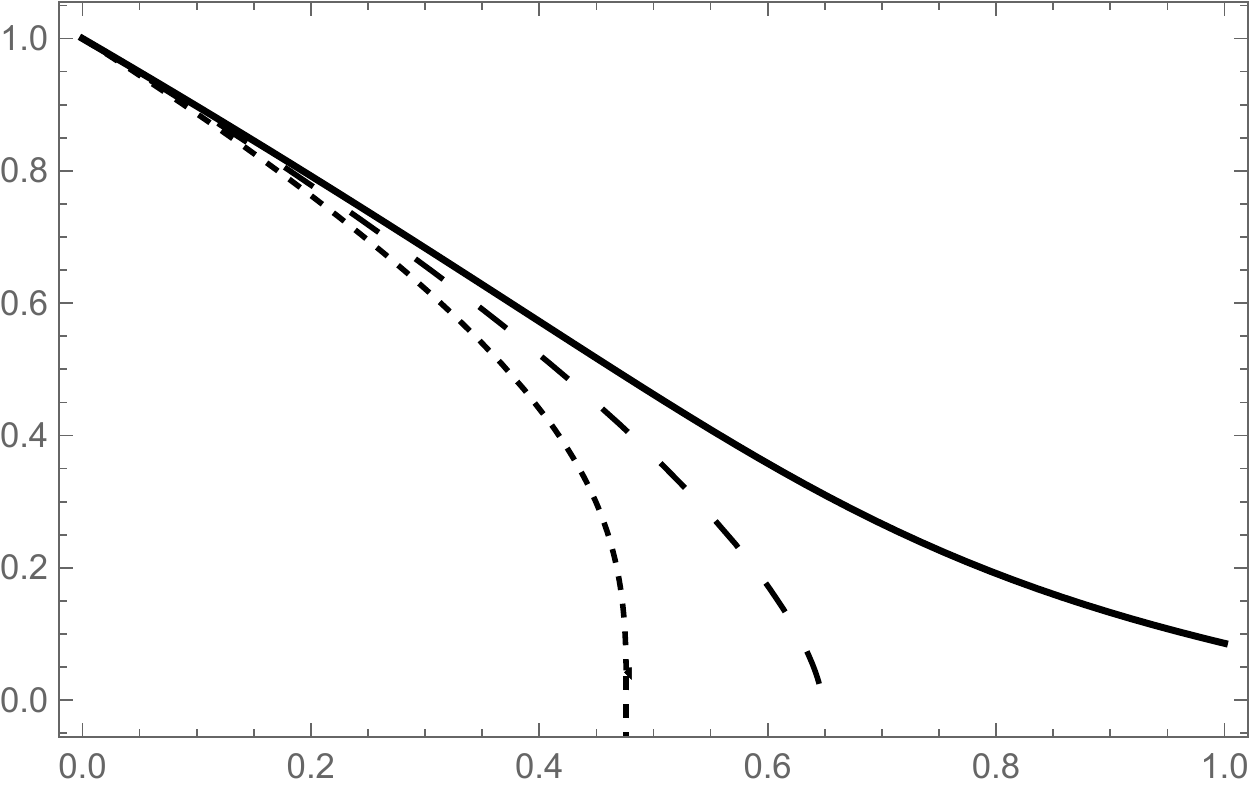}
\put(6,50){$a(t)$}
\put(90,6){$t$}
\end{overpic}
\end{minipage}
\caption{Semi-classical unbound collapse, i.e. $k>0$, with $(+)$ in equation \eqref{din}. Left panel: The function $j(a)=\dot{a}$ is plotted for $\gamma=0$ (large dashed), $\gamma=1$ (short dashed) and $\gamma=-1$ (solid). For the sake of clarity the range of the horizontal axis is $(0,2)$ even if the initial condition is taken as $a(t_i)=1$. Right panel: The scale factor $a(t)$ is plotted for $\gamma=0$ (large dashed), $\gamma=1$ (short dashed) and $\gamma=-1$ (solid). The plots are obtained for $q^3=0.1$ and $k=0.1$.}  \label{plus-k1}
\end{figure*}

\begin{figure*}[ht]
\begin{minipage}{0.49\linewidth}
\begin{overpic}
[width=0.97\linewidth]{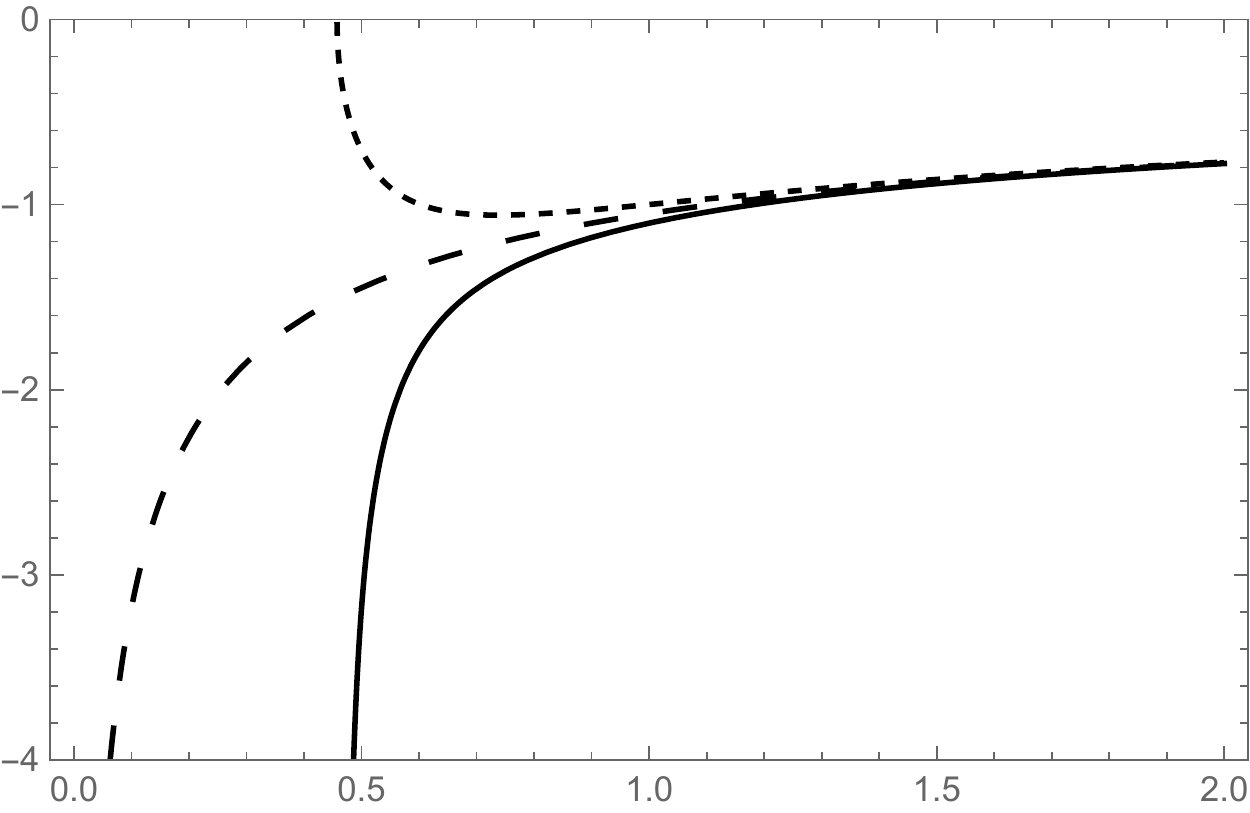}
\put(6,50){$j(a)$}
\put(90,6){$a$}
\end{overpic}
\end{minipage}
\hfill 
\begin{minipage}{0.49\linewidth}
\begin{overpic}
[width=0.97\linewidth]{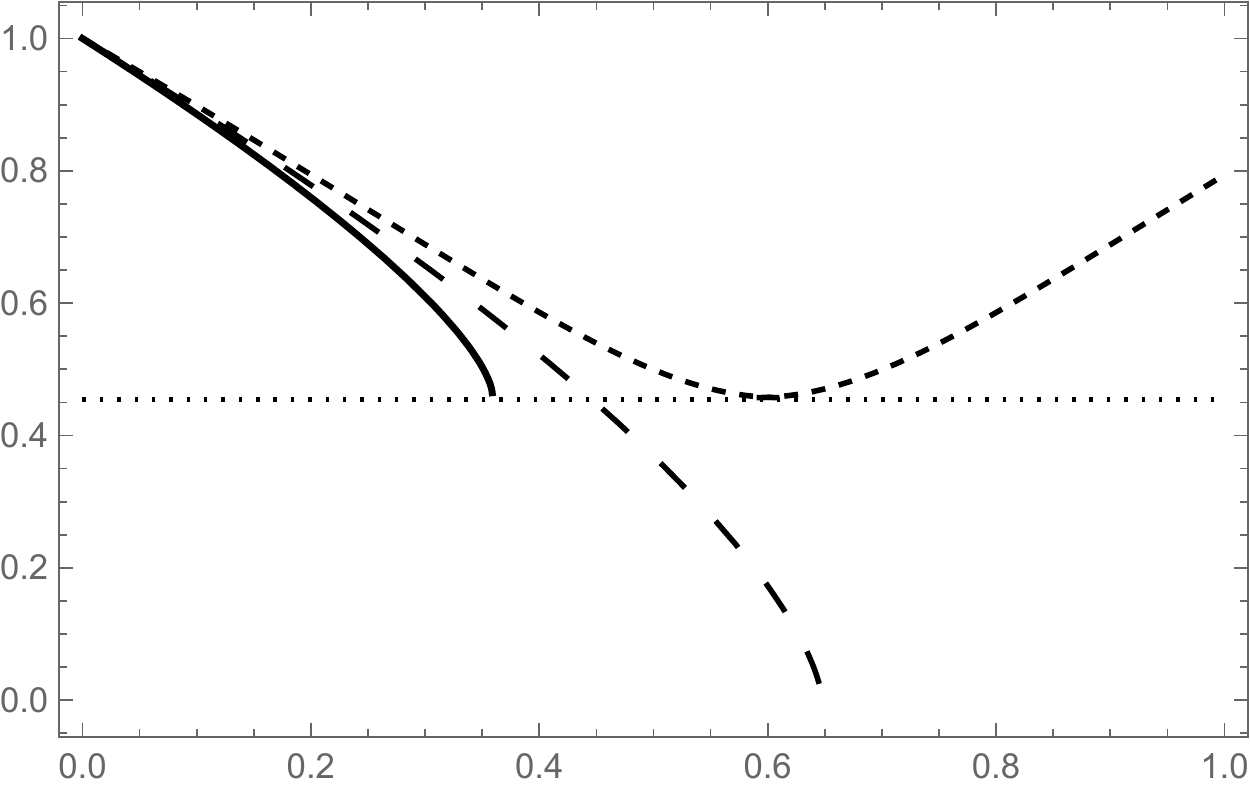}
\put(6,50){$a(t)$}
\put(90,6){$t$}
\end{overpic}
\end{minipage}
\caption{Semi-classical unbound collapse, i.e. $k>0$, with$(-)$ in equation \eqref{din}. Left panel: The function $j(a)=\dot{a}$ is plotted for $\gamma=0$ (large dashed), $\gamma=1$ (short dashed) and $\gamma=-1$ (solid). For the sake of clarity the range of the horizontal axis is $(0,2)$ even if the initial condition is taken as $a(t_i)=1$. Right panel: The scale factor $a(t)$ is plotted for $\gamma=0$ (large dashed), $\gamma=1$ (short dashed) and $\gamma=-1$ (solid). The plots are obtained for $q^3=0.1$ and $k=0.1$.}  \label{minus-k1}
\end{figure*}

\begin{figure*}[tt]
\begin{minipage}{0.49\linewidth}
\begin{overpic}
[width=0.97\linewidth]{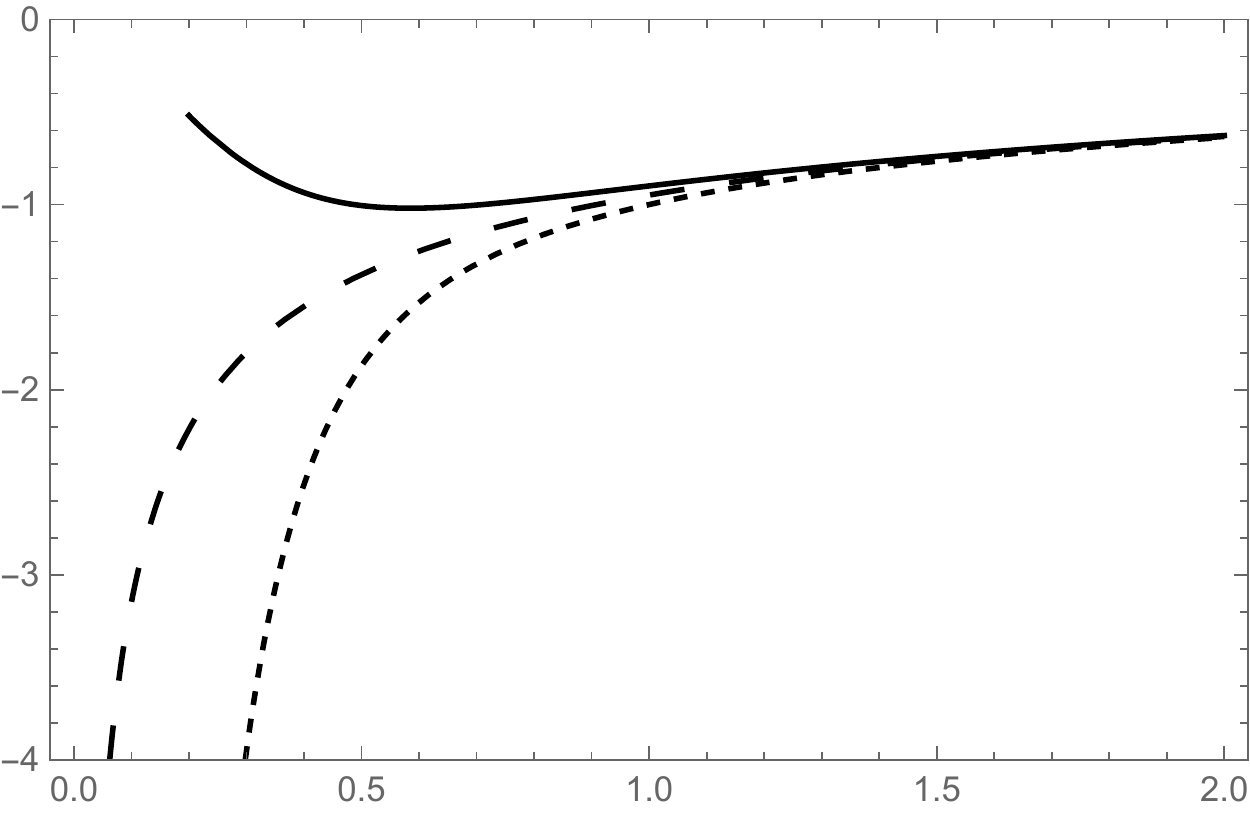}
\put(6,50){$j(a)$}
\put(90,6){$a$}
\end{overpic}
\end{minipage}
\hfill 
\begin{minipage}{0.49\linewidth}
\begin{overpic}
[width=0.97\linewidth]{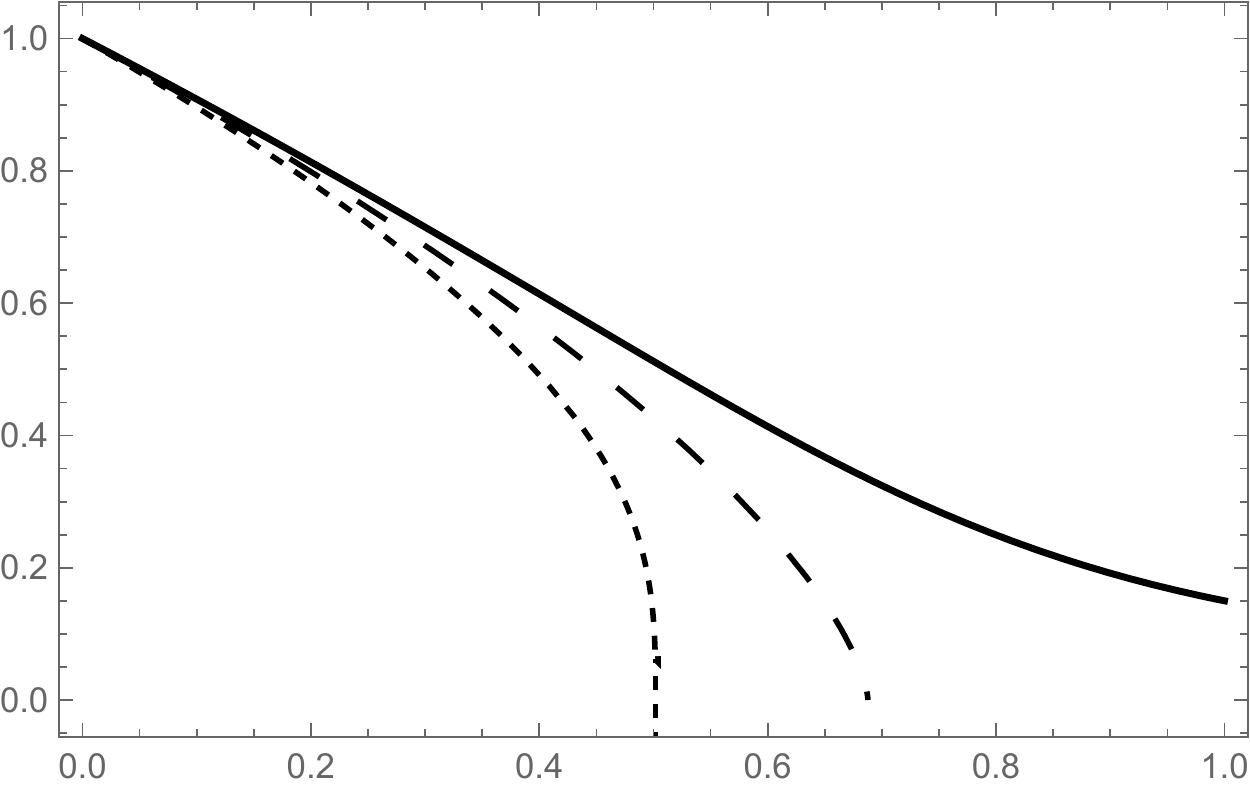}
\put(6,50){$a(t)$}
\put(90,6){$t$}
\end{overpic}
\end{minipage}
\caption{Semi-classical bound collapse, i.e. $k<0$, with $(+)$ in equation \eqref{din}. Left panel: The function $j(a)=\dot{a}$ is plotted for $\gamma=0$ (large dashed), $\gamma=1$ (short dashed) and $\gamma=-1$ (solid). For the sake of clarity the range of the horizontal axis is $(0,2)$ even if the initial condition is taken as $a(t_i)=1$. Right panel: The scale factor $a(t)$ is plotted for $\gamma=0$ (large dashed), $\gamma=1$ (short dashed) and $\gamma=-1$ (solid). The plots are obtained for $q^3=0.1$ and $k=-0.1$.} \label{plus-k-1}
\end{figure*}

\begin{figure*}[ht]
\begin{minipage}{0.49\linewidth}
\begin{overpic}
[width=0.97\linewidth]{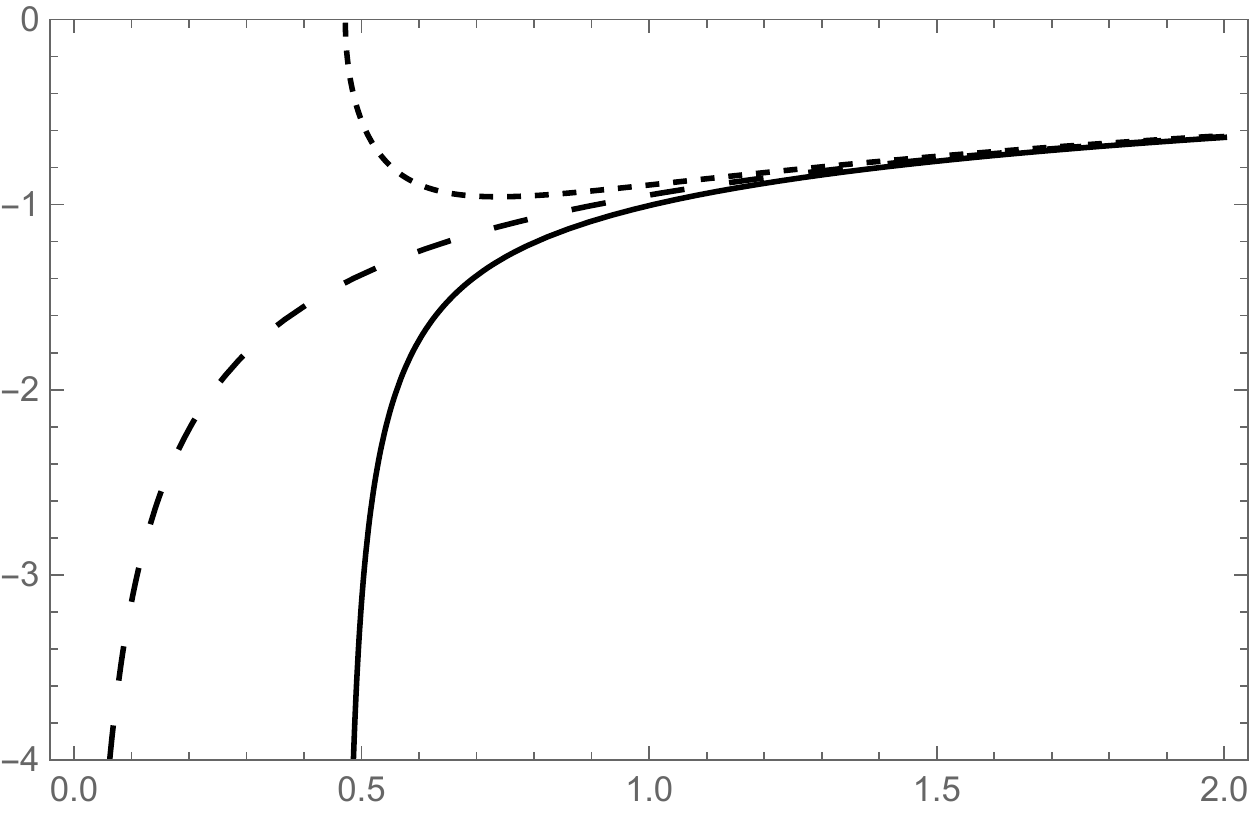}
\put(6,50){$j(a)$}
\put(90,6){$a$}
\end{overpic}
\end{minipage}
\hfill 
\begin{minipage}{0.49\linewidth}
\begin{overpic}
[width=0.97\linewidth]{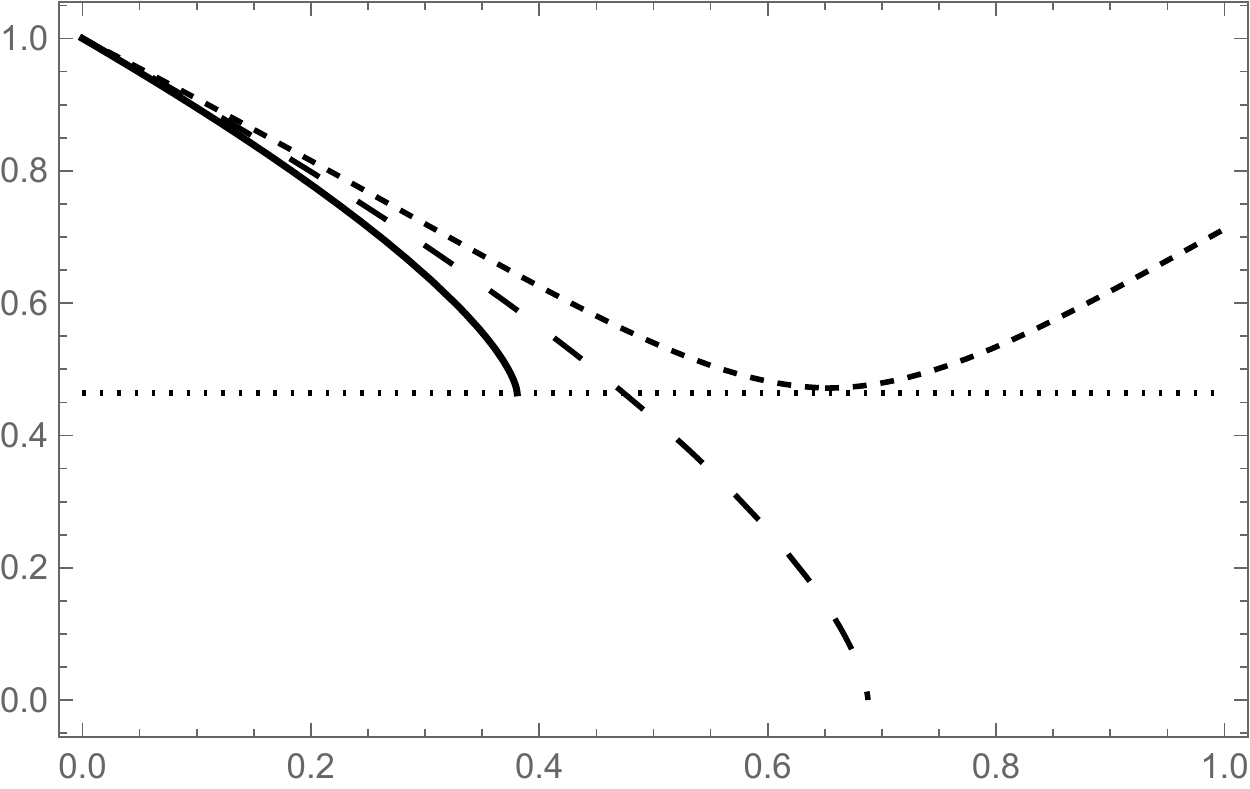}
\put(6,50){$a(t)$}
\put(90,6){$t$}
\end{overpic}
\end{minipage}
\caption{Semi-classical bound collapse, i.e. $k<0$, with $(-)$ in equation \eqref{din}. Left panel: The function $j(a)=\dot{a}$ is plotted for $\gamma=0$ (large dashed), $\gamma=1$ (short dashed) and $\gamma=-1$ (solid). For the sake of clarity the range of the horizontal axis is $(0,2)$ even if the initial condition is taken as $a(t_i)=1$. Right panel: The scale factor $a(t)$ is plotted for $\gamma=0$ (large dashed), $\gamma=1$ (short dashed) and $\gamma=-1$ (solid). The plots are obtained for $q^3=0.1$ and $k=-0.1$.} \label{minus-k-1}
\end{figure*}

These possibilities are illustrated for three possible values of $k=0,\pm0.1$ in Figures \ref{plus-k0}-\ref{minus-k-1}. In Figure \ref{plus-k0} are shown $j(a)$ (left panel) and $a(t)$ (right panel) for marginally bound collapse models with $(+)$ for three values of $\gamma=0,\pm 1$. The scale factor $a$ goes to zero in all three cases but $j$ goes to zero only in the case $\gamma=-1$, which corresponds to the Hayward black hole \cite{bobir}. The corresponding models with $k=0.1$ and $k=-0.1$ are shown in Figure \ref{plus-k1} and Figure \ref{plus-k-1} respectively.

In Figure \ref{minus-k0} are shown $j(a)$ (left panel) and $a(t)$ (right panel) for marginally bound collapse models with $(-)$ for three values of $\gamma=0,\pm 1$. The scale factor $a$ goes to a finite value if $\gamma\neq 0$ and $j$ goes to zero only in the case $\gamma=1$, which corresponds to the bouncing model discussed in \cite{collapse6}. The corresponding models with $k=0.1$ and $k=-0.1$ are shown in Figure \ref{minus-k1} and Figure \ref{minus-k-1} respectively.

Physically, in order to avoid the formation of the singularity, it seems reasonable to look for those models for which $j$ goes to zero, i.e. collapse halts, either in a finite time or asymptotically. Also for collapse to settle to an equilibrium configuration we should require that $\ddot{a}$ goes to zero. Finally we wish the endstate of collapse to be a regular black hole. Notice that $\gamma>0$, for which $j$ goes to zero in the $(-)$ case, implies $\lambda<0$ and thus the corresponding solutions in GR coupled to NLED can not be regular black holes. On the other hand, having set $\beta=1$, we see that $\gamma\leq -1$ (namely $\lambda\geq 3$) satisfies all three criteria and thus the corresponding solutions in GR coupled to NLED are regular black holes that originate as the endstate of collapse.


\subsection{Trapped surfaces}

As matter collapses trapped surfaces may form. The equation that implicitly defines the apparent horizon in the interior cloud is \eqref{apparent-horizon} or
\be \label{ah}
1-\frac{r^2m_{\rm eff}}{a}=0\, ,
\ee 
which in the marginally bound cases reduces to
\be 
1-r^2\dot{a}^2=0 \;,
\ee 
and implicitly gives the apparent horizon curve as 
\be 
r_{\rm ah}(t)=\frac{1}{|\dot{a}|}\;.
\ee 
As mentioned previously the apparent horizon is present only for $r_{\rm ah}(t)\leq r_b$ and when $r_{\rm ah}(t)=r_b$ it crosses the boundary and connects with a corresponding horizon in the exterior. 

If $\dot{a}\rightarrow -\infty$ then $r_{\rm ah}\rightarrow 0$ and  $r_{\rm ah}(t)$ will cross the boundary only once. Therefore we can not have the formation of an inner horizon during collapse. This is the case of the OSD model. On the other hand if $\dot{a}\rightarrow 0$ then $r_{\rm ah}\rightarrow +\infty$ and it may cross the boundary twice thus allowing for the possibility of producing the outer and inner horizons. This is the case of the Hayward regular black hole. Interestingly, the regular black hole is not the only possible option for the exterior if $\dot{a}\rightarrow 0$. In fact the horizon in the exterior may be closed giving rise to a closed trapped surface that exists for a finite time. These cases may be described with a Vaidya or generalized Vaidya exterior.

\begin{figure*}[ttt]
\begin{minipage}{0.49\linewidth}
\center{\includegraphics[width=0.97\linewidth]{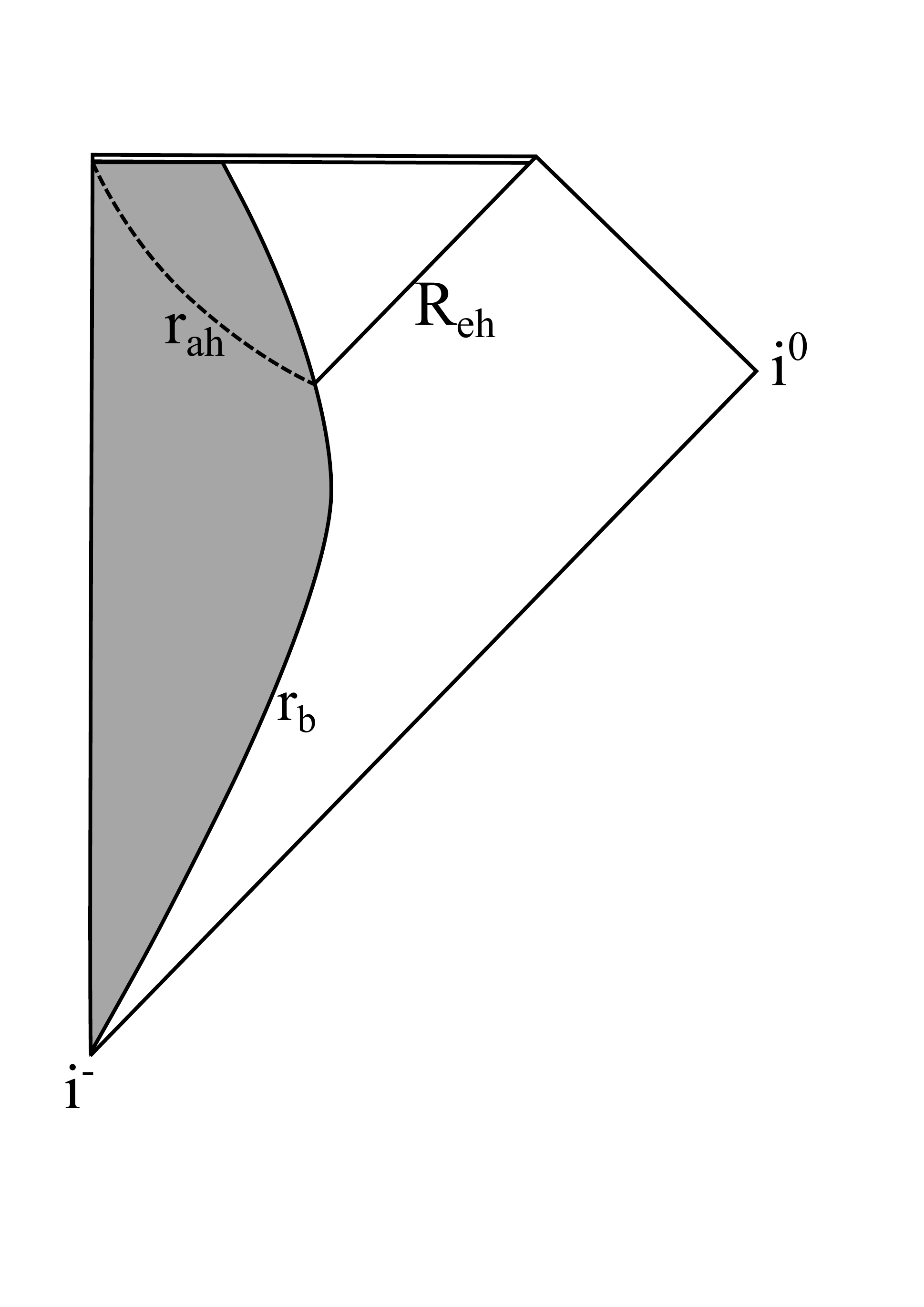}\\ }
\end{minipage}
\hfill 
\begin{minipage}{0.49\linewidth}
\center{\includegraphics[width=0.97\linewidth]{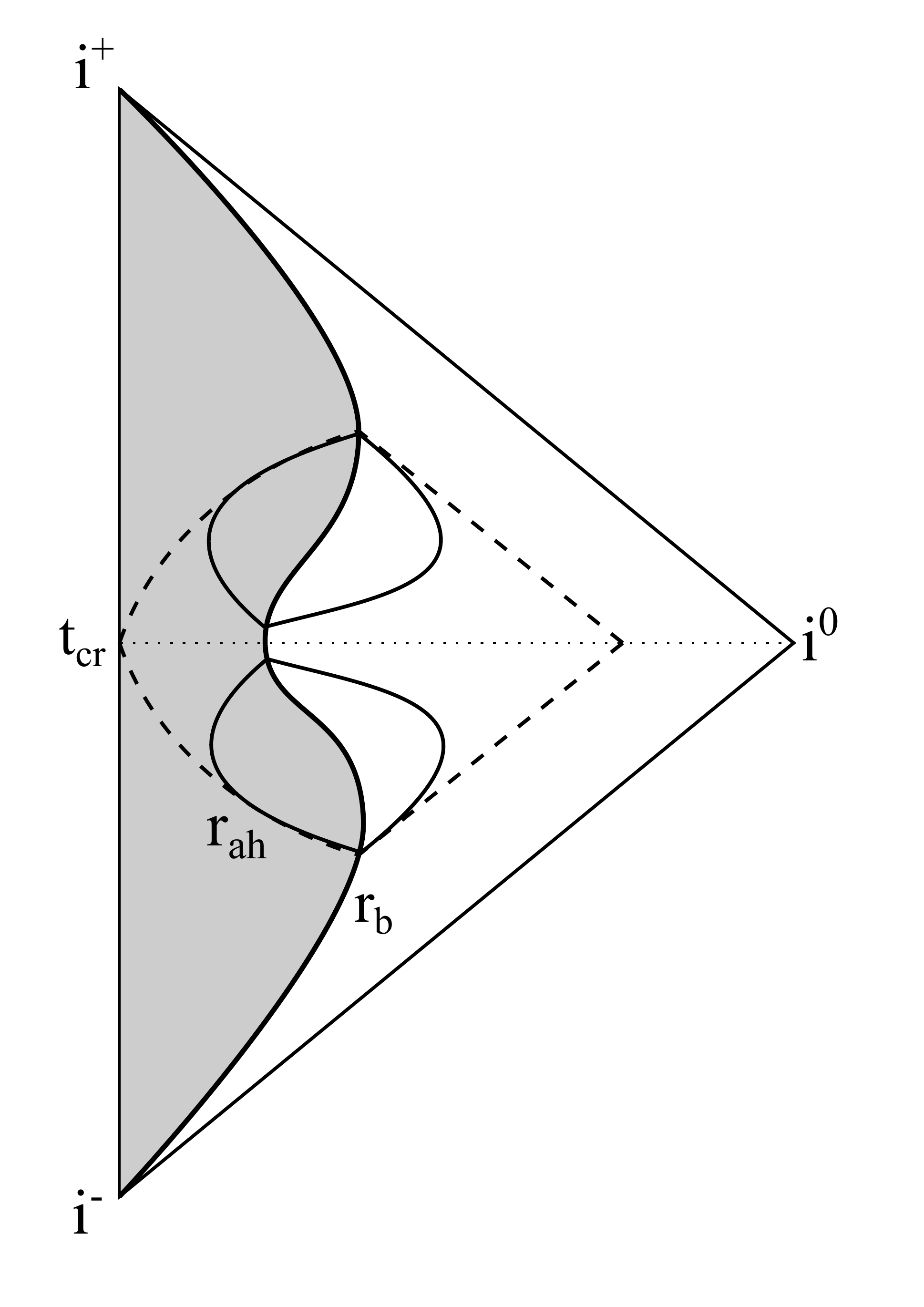}\\ }
\end{minipage}
\caption{Left panel: Penrose diagram for the OSD collapse. The darker region represents the interior of the cloud with boundary $r_b$ while the double horizontal line represents the singularity. The solid line $R_{\rm eh}$ in the exterior is the event horizon while the dashed line in the interior is the apparent horizon. Right panel: Penrose diagram for the LQG inspired collapse model with dynamical Vaidya-like exterior. The darker region represents the interior of the cloud with boundary $r_b$. Collapse turns into expansion at the time $t_{\rm cr}$ and the spacetime is regular everywhere. The solid closed lines represent the trapped regions before and after the bounce while the dashed lines represent the event horizon and apparent horizon of the OSD case. The expanding solution for $t>t_{\rm cr}$ is given by the time reversal of the collapsing one.}\label{Penrose-1}
\end{figure*}

\begin{figure*}[ttt]
\begin{minipage}{0.49\linewidth}
\center{\includegraphics[width=0.97\linewidth]{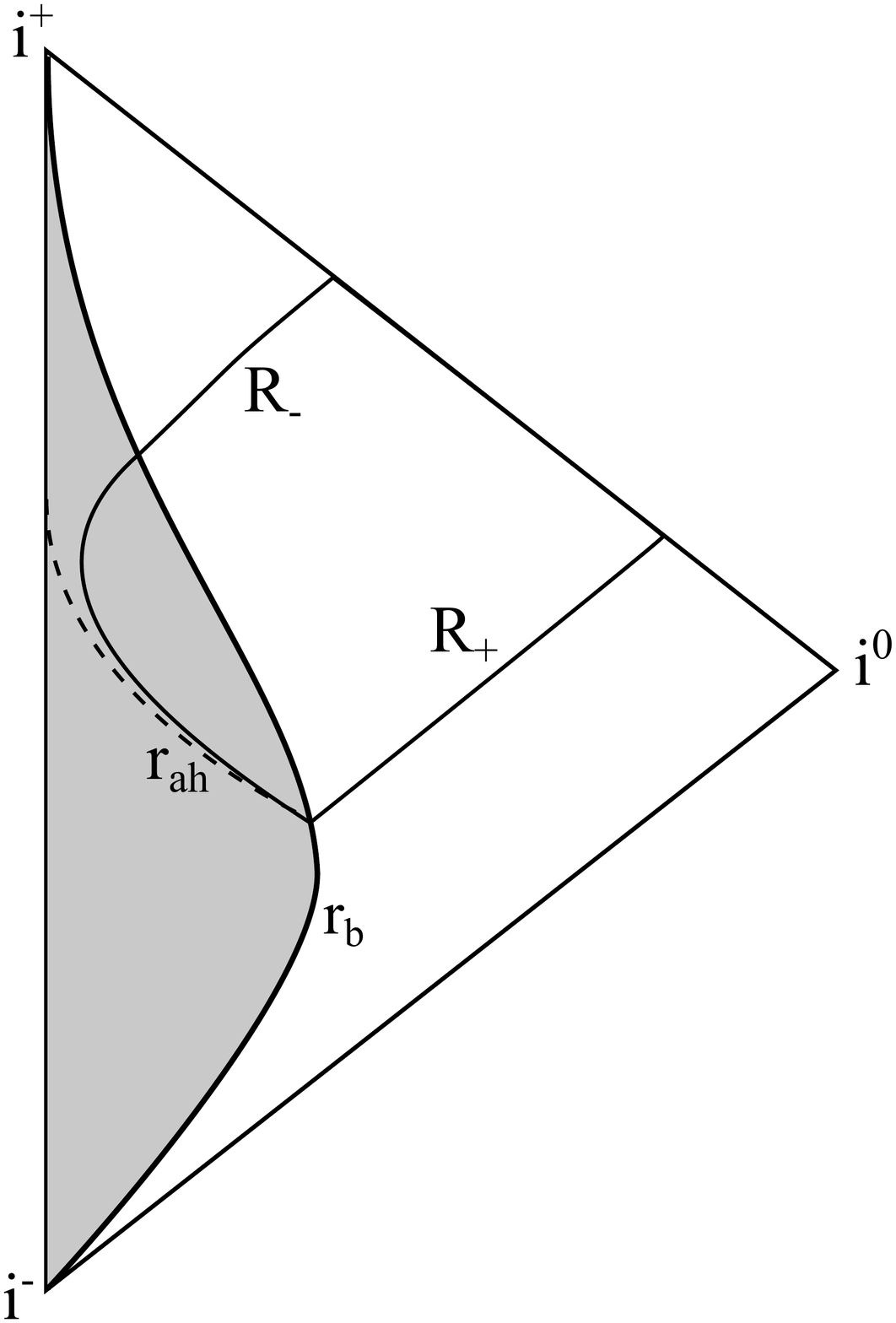}\\ }
\end{minipage}
\hfill 
\begin{minipage}{0.49\linewidth}
\center{\includegraphics[width=0.97\linewidth]{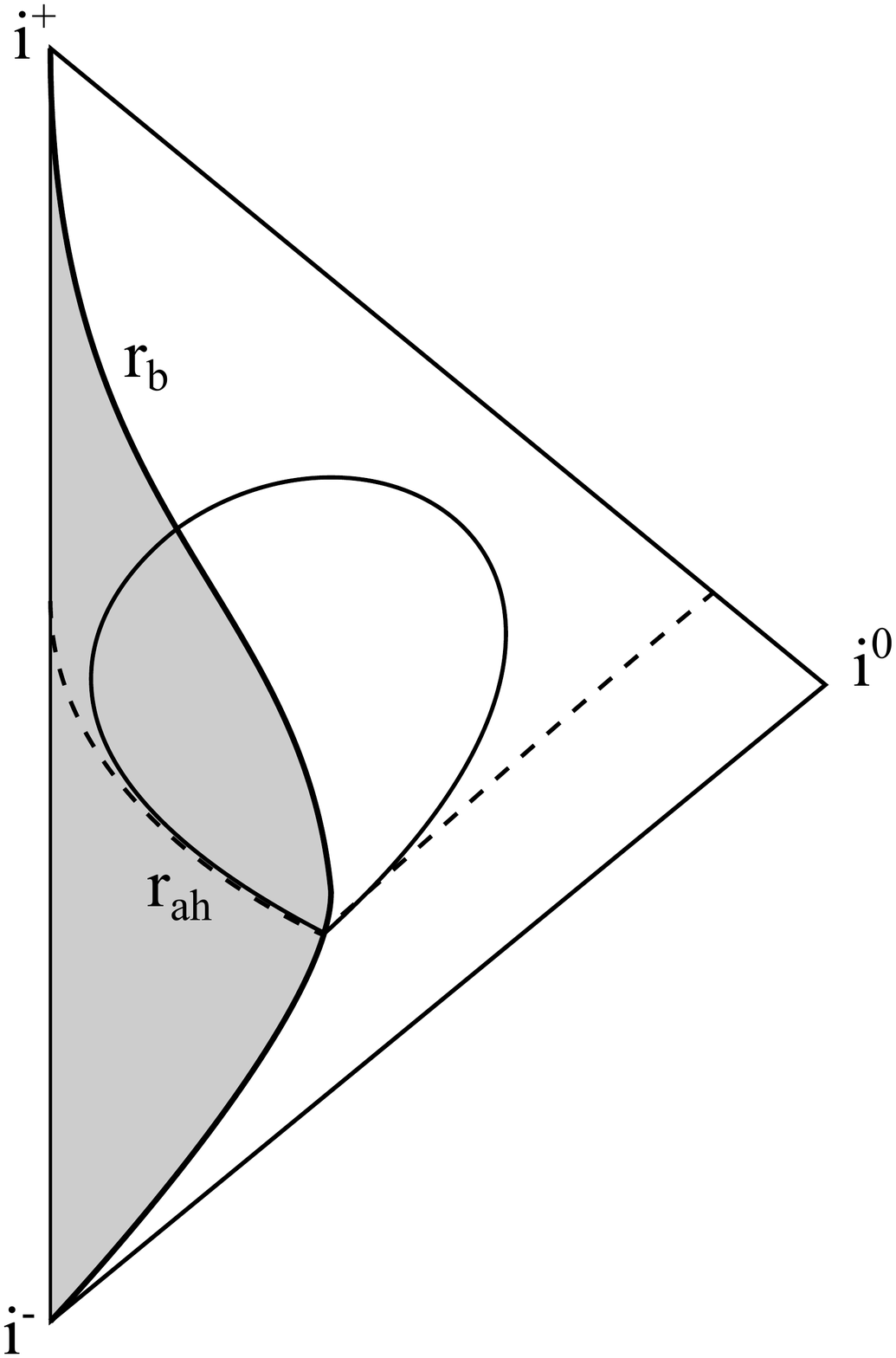}\\ }
\end{minipage}
\caption{Left panel: Penrose diagram for the collapse model leading to the Hayward black hole. The darker region represents the interior of the cloud with boundary $r_b$ while solid lines represent the horizons, namely the apparent horizon $r_{\rm ah}$ in the interior and the inner $R_-$ and outer $R_+$ horizons in the exterior. The dashed line in the interior represents the apparent horizon of the OSD case. Right panel: Penrose diagram for collapse model leading to a horizonless remnant. The darker region represents the interior of the cloud with boundary $r_b$. The solid closed line represents the trapped region while the dashed lines represent the event horizon and apparent horizon of the OSD case.} \label{Penrose-2}
\end{figure*}

To summarize we have four possible scenarios:
\begin{enumerate}
    \item Only one horizon forms and the singularity forms at the end of collapse. This is the case of OSD collapse. This case is shown in the left panel of Figure \ref{Penrose-1}.
    \item A closed trapped surface forms before the bounce and the singularity is averted. If the expansion after the bounce is the time reversal of the collapse case a second closed trapped surface will form, this time describing a white hole instead of a black hole. This is the case of the LQG inspired model \cite{collapse6}. This case is shown in the right panel of Figure \ref{Penrose-1}.
    \item An inner and an outer horizon form as matter settles asymptotically and the singularity is averted. The outer spacetime is described by a regular black hole. This is the case discussed in \cite{bobir}. This case is shown in the left panel of Figure \ref{Penrose-2}.
    \item A closed trapped surface forms as matter settles asymptotically and the singularity is averted. The outer spacetime is described by an horizonless compact remnant.  This case is shown in the right panel of Figure \ref{Penrose-2}.
\end{enumerate}

\subsection{Example: Collapse and bounce in LQG}

\begin{figure*}[ttt]
\begin{minipage}{0.49\linewidth}
\begin{overpic}
[width=0.97\linewidth]{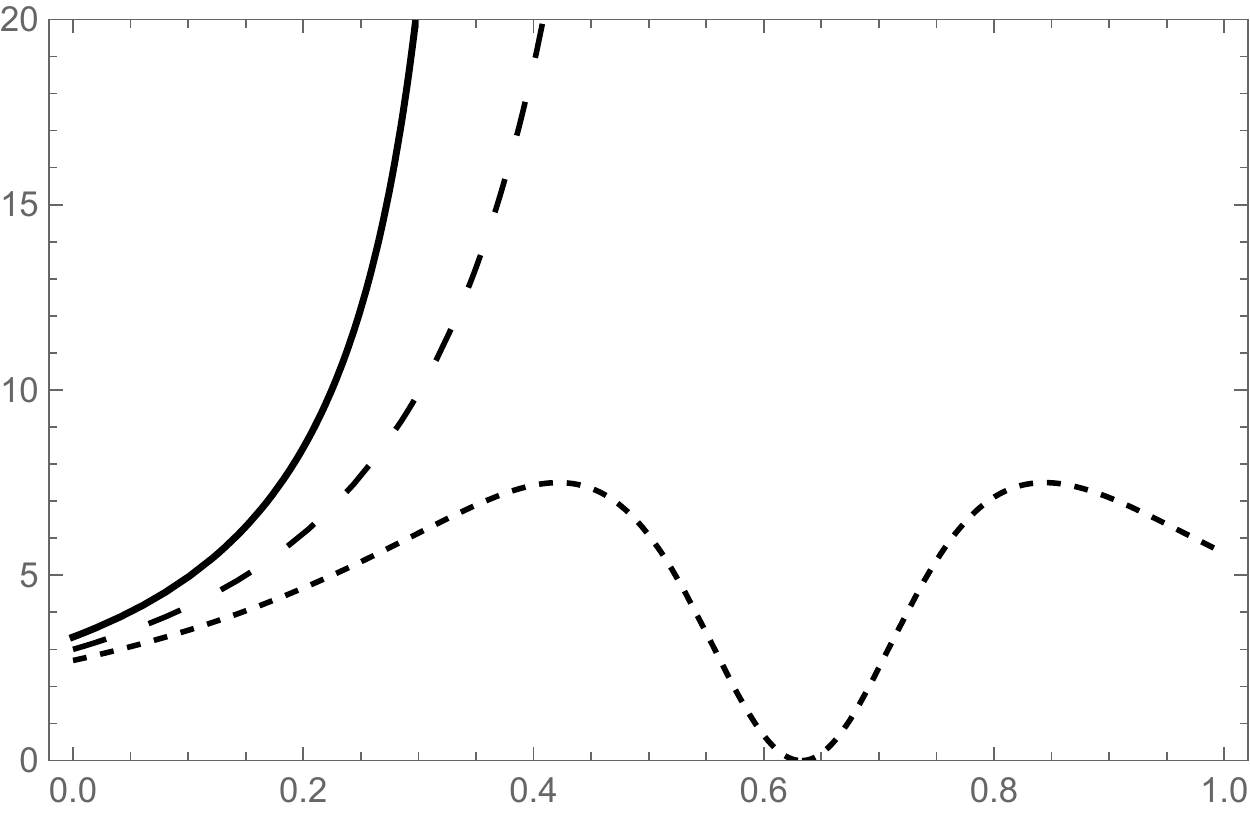}
\put(6,50){$\epsilon_{\rm eff}(t)$}
\put(90,6){$t$}
\end{overpic}
\end{minipage}
\hfill 
\begin{minipage}{0.49\linewidth}
\begin{overpic}
[width=0.97\linewidth]{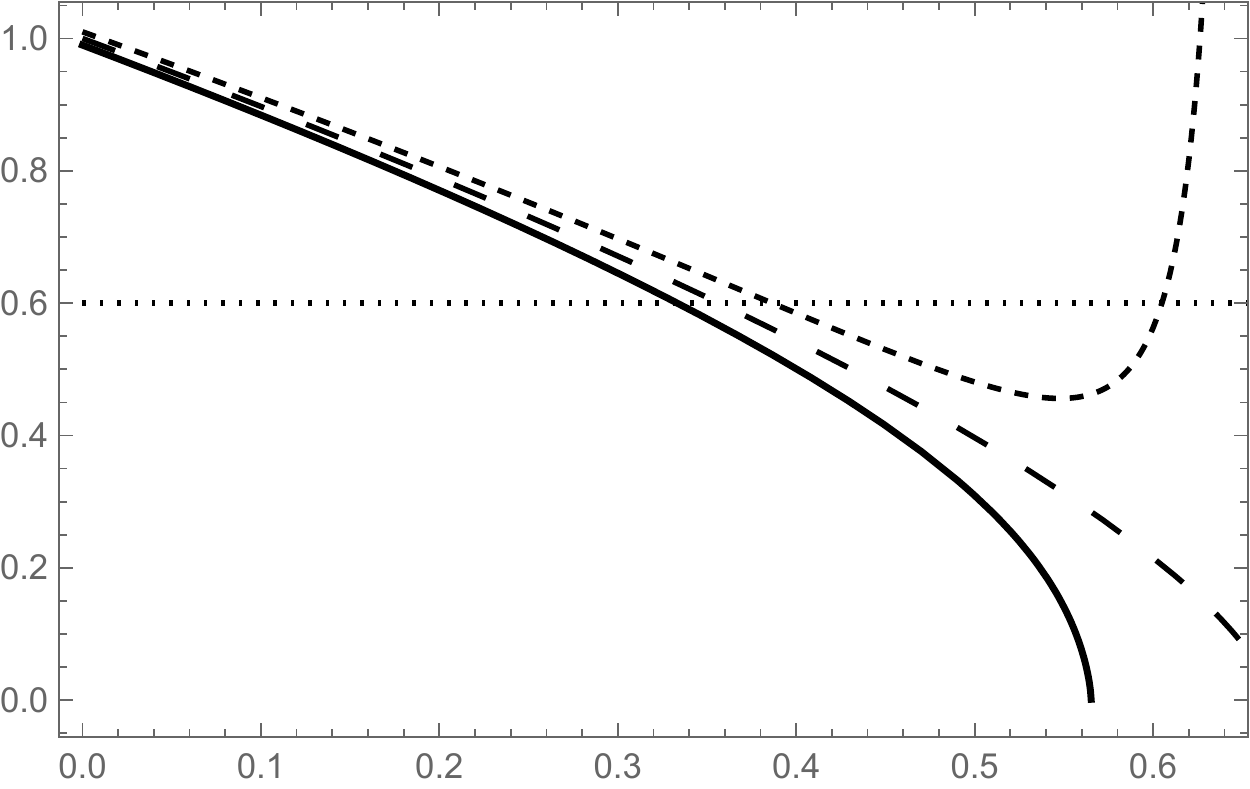}
\put(6,50){$r_{\rm ah}(t)$}
\put(90,6){$t$}
\end{overpic}
\end{minipage}
\caption{Semi-classical marginally bound collapse, i.e. $k=0$, with $(-)$ in equation \eqref{din}. Left panel: The effective density $\epsilon_{\rm eff}(t)$ is plotted for $\gamma=0$ (large dashed), $\gamma=1$ (short dashed) and $\gamma=-1$ (solid). The effective density is obtained for $q^3=0.1$. Right panel: The apparent horizon $r_{\rm ah}(t)$ is plotted for $\gamma=0$ (large dashed), $\gamma=1$ (short dashed) and $\gamma=-1$ (solid). The boundary at $r_b=0.6$ is represented by the horizontal dotted line. The case $\gamma=0$ corresponds to OSD collapse where $\epsilon_{\rm eff}=\epsilon$, while $\gamma=1$ corresponds to the LQG inspired collapse.  The apparent horizon is obtained for $q^3=0.01$.}\label{example1}
\end{figure*}

The LQG inspired model discussed in \cite{collapse6} bounces turning collapse into expansion in a finite time. This model is obtained in the above formalism by taking the $(-)$ case, $k=0$ and $\gamma=1$, $\beta=1$ (i.e. $\lambda=-3$, $\kappa=3$) in equation \eqref{semi-eom}. The effective energy density in this model is given by equation \eqref{LQG-epsilon} and relates to the effective Misner-Sharp mass via
\be 
\epsilon_{\rm eff}=\frac{3m_{\rm eff}}{a^3} \, ,
\ee
and is shown in the left panel of Figure \ref{example1}. The resulting effective pressure is given by
\be 
p_{\rm eff}=-\frac{\dot{m}_{\rm eff}}{a^2\dot{a}}=-\frac{\epsilon^2}{\epsilon_{\rm cr}} \; .
\ee 
Notice that the effective pressure is negative and vanishes for $\epsilon_{\rm cr}\rightarrow +\infty$ which corresponds to the OSD case.
The effective Misner-Sharp mass $m_{\rm eff}$ still obeys equation \eqref{misner-3}, which for homogeneous dust gives equation \eqref{LQG-adot}, which can be written as
\be 
m_{\rm eff}=m_0\left(1-\frac{a_{\rm cr}^3}{a^3}\right)=a(\dot{a}^2-k) \, .
\ee
Notice that to the critical density parameter $\epsilon_{\rm cr}$ corresponds a critical scale for collapse $a_{\rm cr}$ from $\epsilon_{\rm cr}=3m_0/a_{\rm cr}^3$.
To retrieve the OSD model we must consider the limit $\epsilon_{\rm cr}\rightarrow +\infty$, which corresponds to $a_{\rm cr}\rightarrow 0$. The solution for the marginally bound case is then easily obtained as
\be 
a(t)=\left(a_{\rm cr}^3+\left(\sqrt{1-a_{\rm cr}^3}-\frac{3}{2}\sqrt{m_0}t\right)^2\right)^{1/3} \; ,
\ee 
which reduces to equation \eqref{a(t)} for $a_{\rm cr}^3=0$. The effective density and apparent horizon for this model are shown as the dashed lines in the left and right panels of Figure \ref{example1}, respectively.

In this model the scale factor reaches the minimum value $a_{\rm cr}$ in a finite co-moving time $t_{\rm cr}$ and then collapse turns into expansion as the matter bounces. From the apparent horizon equation \eqref{ah} we see that $r_{\rm ah}=r_b$ has two solutions for $t<t_{\rm cr}$ showing that the trapping horizon crosses the boundary twice. The behavior of the interior solution does not tell us anything about the exterior. In fact we have two possibilities for the exterior that satisfy the matching conditions. The first possibility, as suggested in \cite{collapse6}, is that a black hole forms during collapse but the horizon eventually disappears and as the matter bounces at $t=t_{\rm cr}$ the central region is not covered by the horizon (see the right panel in Figure \ref{example1}). This scenario is consistent with the fact that at $t_{\rm cr}$ the effective Misner-Sharp mass vanishes suggesting that the spacetime at the time slice $t=t_{\rm cr}$ is flat. This scenario is not matched to a static black hole such as the ones described by GR coupled to NLED because at $t=t_{\rm cr}$ there must be no horizon anywhere in the spacetime, while we have seen that black holes coupled to NLED must have at least two horizons. In fact it can be shown that the LQG inspired collapse model can be matched to an exterior with variable mass as described by the Vaidya metric \cite{collapse8}. The second possibility, also discussed in \cite{collapse8}, is that the interior is matched to a static exterior of the form \eqref{metric-rbh}. In this case the two roots of $r_{\rm ah}=r_b$ match with the inner and outer horizons of the exterior. However, this scenario does not produce a regular black hole in NLED, as it can also be seen from the fact that $\lambda=-3$. 

We may still ask what exterior spacetime would result from the corresponding NLED Lagrangian and consider its properties. It is easy to see that the NLED Lagrangian for this model is
\be 
\mathcal{L}_{\rm NLED}(\mathrm{F})=-12\sqrt{\alpha}\mathrm{F}^{3/2} \, ,
\ee 
and $f(R)$ is given by equation \eqref{M} with $q_*=-a_{\rm cr}$.
The Kretschmann scalar for this metric diverges for $R\rightarrow 0$ as
\be 
\mathcal{K}=\frac{48M^2}{R^{12}}(39q_*^6-10q_*^3R^3+R^6) \, .
\ee 
Nevertheless, looking the equation for the radial infall of a particle in this spacetime we can easily see that $R=0$ can not be reached and a test particle on a radial ingoing trajectory must bounce. In fact the equation of motion for a test particle of energy per unit mass $E$ falling radially along a trajectory $R(\tau)$ is
\be 
E^2-1=\dot{R}^2-\frac{2M_0}{R}+\frac{2M_0q_*^3}{R^4}\;, 
\ee 
which in the case of a particle with zero initial velocity at spatial infinity, i.e. with $E=1$, analogous to the marginally bound collapse, has a turning point at $R=q_*$.

\subsection{Example: Collapse to the Hayward black hole}

\begin{figure*}[ttt]
\begin{minipage}{0.49\linewidth}
\begin{overpic}
[width=0.97\linewidth]{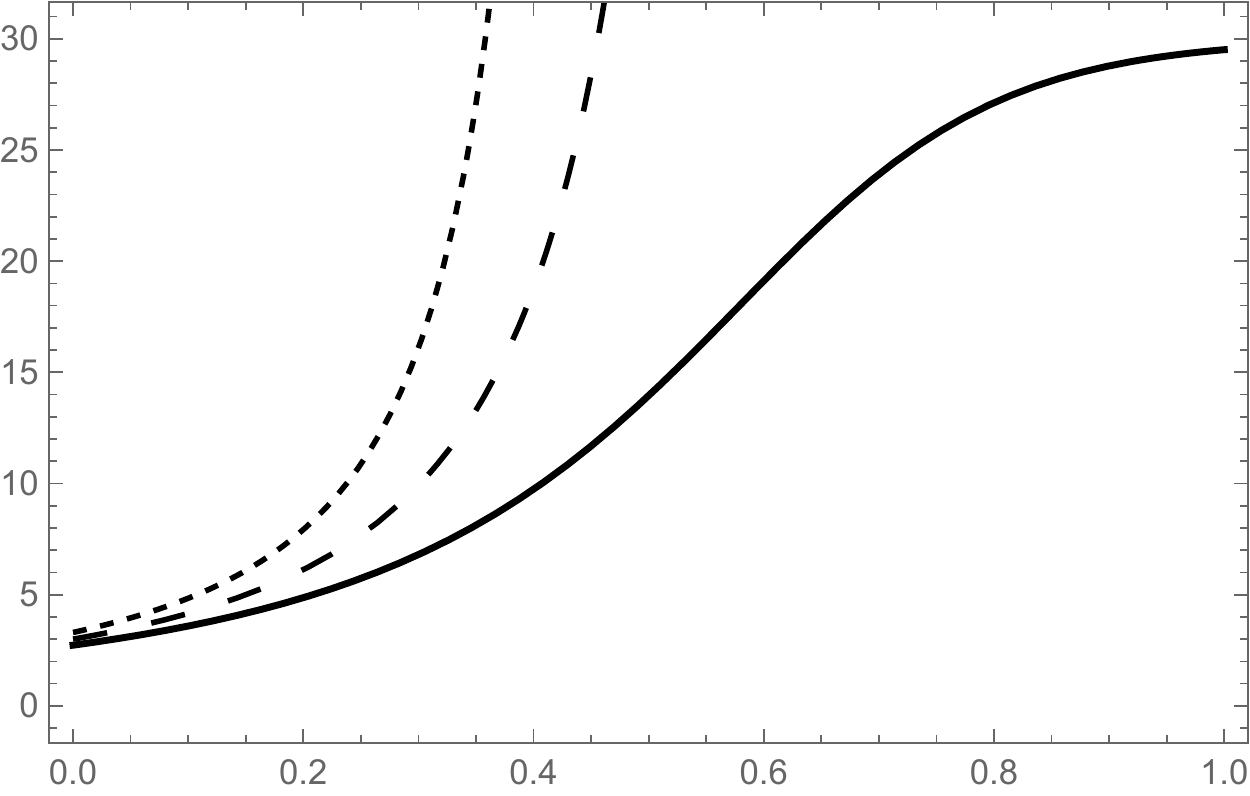}
\put(6,50){$\epsilon_{\rm eff}(t)$}
\put(90,6){$t$}
\end{overpic}
\end{minipage}
\hfill 
\begin{minipage}{0.49\linewidth}
\begin{overpic}
[width=0.97\linewidth]{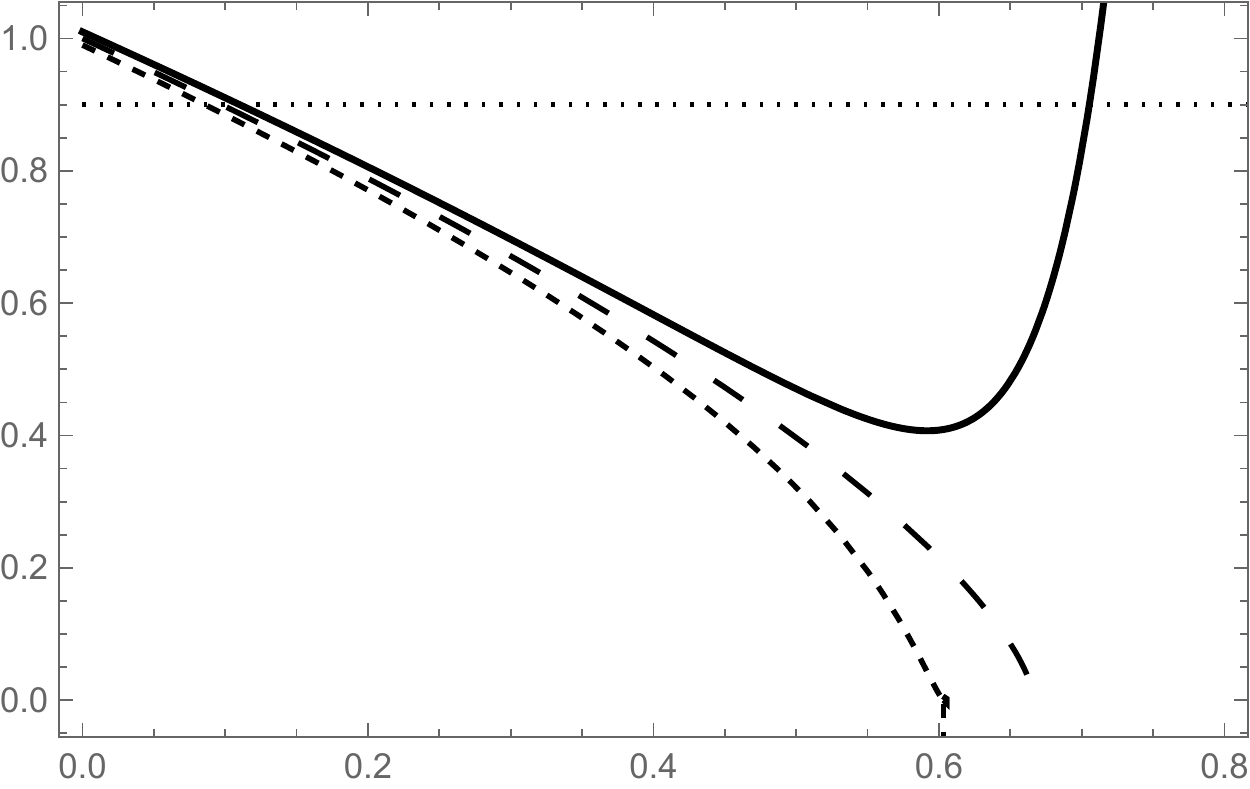}
\put(6,50){$r_{\rm ah}(t)$}
\put(90,6){$t$}
\end{overpic}
\end{minipage}
\caption{Semi-classical marginally bound collapse, i.e. $k=0$, with $(+)$ in equation \eqref{din}. Left panel: The effective density $\epsilon_{\rm eff}(t)$ is plotted for $\gamma=0$ (large dashed), $\gamma=1$ (short dashed) and $\gamma=-1$ (solid). The effective density is obtained for $q^3=0.1$. Right panel: The apparent horizon $r_{\rm ah}(t)$ is plotted for $\gamma=0$ (large dashed), $\gamma=1$ (short dashed) and $\gamma=-1$ (solid). The boundary at $r_b=0.8$  is represented by the horizontal dotted line.. The case $\gamma=0$ corresponds to OSD collapse where $\epsilon_{\rm eff}=\epsilon$, while $\gamma=-1$ corresponds to the NLED collapse leading to the Hayward regular black hole. The apparent horizon is obtained for $q^3=0.01$.}\label{example2}
\end{figure*}

For black hole solutions in GR coupled to NLED we retrieve the Hayward regular black hole \cite{bh3} by taking $\lambda=\kappa=3$, which corresponds to $\gamma=-1$, $\beta=1$ in the $(+)$ case in equation \eqref{semi-eom}. We can then investigate the corresponding semi-classical marginally bound dust collapse, which was considered in \cite{bobir}. The effective density is
\be 
\epsilon_{\rm eff}=\epsilon\left(1-\frac{\epsilon}{\epsilon_{\rm cr}+\epsilon}\right)\, .
\ee 
Notice that for small densities this collapse model behaves in the same manner as the LQG inspired one. In fact expanding $\epsilon_{\rm eff}$ for $\epsilon<<\epsilon_{\rm cr}$ we get
\be 
\epsilon_{\rm eff}=\sum_{n=0}^{\infty}\epsilon\frac{(-\epsilon)^n}{\epsilon_{\rm cr}^n}\; ,
\ee 
which stopping at $n=1$ is the same as equation \eqref{LQG-epsilon}.
However, while for the LQG inspired model all higher order terms vanish, in this case they are non zero. As a consequence for $\epsilon=\epsilon_{\rm cr}$ we have that $\epsilon_{\rm eff}=\epsilon_{\rm cr}/2$, while in the LQG case the effective density vanishes at $\epsilon_{\rm cr}$. Also, for this case the effective density is always positive and it reaches the limiting value $\epsilon_{\rm cr}$ asymptotically for $\epsilon\rightarrow +\infty$. As a consequence the scale factor goes to zero asymptotically, as shown in Figure \ref{plus-k0}. The effective density for this model is shown in the right panel of Figure \ref{example2}.

The effective pressure for this model is
\be 
p_{\rm eff}=-\frac{\epsilon^2}{\epsilon_{\rm cr}+\epsilon}\left(1-\frac{\epsilon}{\epsilon_{\rm cr}+\epsilon}\right) \; ,
\ee 
and the equation of motion for the scale factor is
\be 
\dot{a}=-\frac{\sqrt{m_0}a}{\sqrt{a^3+q^3}}\, .
\ee 
In this case $a\rightarrow 0$ as $t$ goes to infinity and the effective density remains finite and goes to the maximum value $\epsilon_{\rm eff}\rightarrow 3m_0/q^3$ asymptotically.
Notice that $\dot{a}\rightarrow 0$ so, if trapped surfaces develop, the apparent horizon equation \eqref{ah} must have two times at which $r_{\rm ah}=r_b$, thus forming the outer and inner horizons, as shown in the right panel of Figure \ref{example2}. Also notice that since $\ddot{a}\rightarrow 0$ collapse must halt asymptotically and does not bounce, leaving the Hayward black hole solution as a late time remnant of collapse. The effective density and apparent horizon for this model are shown as the solid lines in the left and right panels of Figure \ref{example2}, respectively.

Collapse to the Bardeen black hole \cite{bh1,bardeen2} can be obtained in a similar manner for $\lambda=3$ and $\kappa=2$. Other black holes obtained in GR coupled to NLED, such as the ones described in \cite{nled6}, may also be similarly translated into collapse models.

\section{Conclusions}\label{sec5}

We know that General Relativity is an incomplete theory. The generic presence of singularities in physically viable solutions of Einstein's equations suggests that the theory needs to be replaced by another, better suited for the description of gravity at large curvature.  
At present we do not possess such a theory and GR remains the one that best reproduces observations while keeping a simple and beautiful mathematical framework \cite{will}. 
Also we can not tell for sure at what scales (energy, density, size, etc.) the effects of such a new theory would become relevant \cite{zen}. If GR holds unaffected until the Planck scale we may expect it to be replaced by a theory of quantum gravity. Nevertheless there is also the possibility that GR fails before the Planck scale and an alternative classical theory is needed to bridge the gap between GR and quantum gravity.
Therefore, in order to investigate the effects that the new theory of gravity would have in the universe we need to look for models of observable phenomena where the curvature becomes large. Then black holes and gravitational collapse are the ideal candidates.

We have considered here semi-classical models of gravitational collapse. Such models are typically used to describe classically the effects arising at Planck scale from a theory of quantum gravity. However the same formalism may be used also to describe corrections to GR coming from alternative theories or the coupling of GR to some other field, such as a theory of electrodynamics. It is well known that the static black hole solution produced by the coupling of GR with Maxwell's electrodynamics is the Reissner-Nordstrom solution. It is also well known that, under some conditions, from the coupling of GR to a theory of non-linear electrodynamics one may obtain regular black hole solutions such as Hayward's \cite{bh3} or Bardeen's \cite{bh1}. Here we have shown how to obtain such solutions and more from a semi-classical description of gravitational collapse of homogeneous dust.

The study of the properties of such solutions and their possible observational signatures in astrophysical scenarios may prove to be a valuable guide towards our understanding of the limits of GR and the features that a new theory of gravity must posses.


\begin{acknowledgement}
DM acknowledges support from Nazarbayev University Faculty Development Competitive Research Grant No. 11022021FD2926. 
\end{acknowledgement}




\begin{thebibliography}{99}

\bibitem{LQC2} A. Ashtekar, T. Pawlowski and P. Singh, Phys. Rev. D {\bf 73}, 124038 (2006).
\bibitem{LQC3} A. Ashtekar, T. Pawlowski and P. Singh, Phys. Rev. D {\bf 74}, 084003 (2006).
\bibitem{nled1} E. Ay\'on-Beato and A. Garcia, 
Phys. Rev. Lett. {\bf 80}, 5056 (1998).
\bibitem{nled6} E. Ay\'on-Beato and A. Garcia, 
Phys. Lett. B {\bf 464}, 25 (1999).
\bibitem{bardeen2} E. Ay\'on-Beato and A. Garcia, 
Phys. Lett. B {\bf 493}, 149 (2000).
\bibitem{bh6} C. Bambi and L. Modesto, 
Phys. Lett. B {\bf 721}, 329 (2013).
\bibitem{collapse6} C. Bambi, D. Malafarina and L. Modesto, 
Phys. Rev. D {\bf 88}, 044009 (2013).
\bibitem{visser-ec2} C. Barcel\`o and M. Visser, 
Int. J. Mod. Phys. D {\bf 11}, 1553 (2002).
\bibitem{collapse5} C. Barcel\`o, S. Liberati, S. Sonego and M. Visser, 
Phys. Rev. D {\bf 77}, 044032 (2008).
\bibitem{bh1} J.M. Bardeen, in: Conference Proceedings of GR5, Tbilisi, USSR, 174 (1968).
\bibitem{collapse7} F. Ben\`itez, R. Gambini, L. Lehner, S. Liebling and J. Pullin,
Phys. Rev. Lett. {\bf 124}, 071301 (2020).
\bibitem{bojo} M. Bojowald, T. Harada and R. Tibrewala,
Phys. Rev. D {\bf 78}, 064057 (2008).
\bibitem{collapse2} M. Bojowald, 
Phys. Rev. Lett. {\bf 86}, 5227 (2001).
\bibitem{bh2} M. Bojowald, 
Phys. Rev. Lett. {\bf 95}, 061301 (2005).
\bibitem{collapse3} M. Bojowald, R. Goswami, R. Maartens and P. Singh,
Phys. Rev. Lett. {\bf 95}, 091302 (2005).
\bibitem{B} H. Bondi, Mon. Not. Astron. Soc. {\bf 107}, 343 (1947).
\bibitem{horizon2} I. Booth,
Can. Jour. Phys. {\bf 83}, 1073 (2005).
\bibitem{bh9} S. Brahma, C.-Y. Chen and D.-H. Yeom,
Phys. Rev. Lett. {\bf 126}, 181301 (2021).
\bibitem{nled2} K. A. Bronnikov, Phys. Rev. Lett. {\bf 85}, 4641 (2000).
\bibitem{nled3} K. A. Bronnikov, Phys. Rev. D {\bf 63}, 044005 (2001).
\bibitem{nled7} K. A. Bronnikov, Phys. Rev. D {\bf 96}, 128501 (2017).
\bibitem{chandra-book} S. Chandrasekhar, {\em An Introduction to the Study of Stellar Structure}, The University of Chicago press (Chicago, USA, 1939).
\bibitem{Datt} S. Datt, Z. Phys. {\bf 108}, 314 (1938).
\bibitem{Dymnikova} I. Dymnikova  Class. Quantum Grav. {\bf 19}, 725 (2002).
\bibitem{bh10} I. Dymnikova and M. Khlopov,
Int. J. Mod. Phys. D {\bf 24}, 1545002 (2015). 
\bibitem{nled4} Z.-Y. Fan and X. Wang, 
Phys. Rev. D {\bf 94}, 124027 (2016).
\bibitem{matching2} F. Fayos, X. Jaen, E. Llanta and J. M. M. Senovilla, Phys. Rev. D {\bf 45}, 2732 (1992).
\bibitem{matching3} F. Fayos, J. M. M. Senovilla and R. Torres, Phys. Rev. D {\bf 54}, 4862 (1996).
\bibitem{horizon} D. Finkelstein,
Phys. Rev. {\bf 110}, 965 (1958).
\bibitem{bh7} V. P. Frolov, 
Phys. Rev. D {\bf 94}, 104056 (2016).
\bibitem{collapse1} V. P. Frolov and G.A. Vilkovisky,  
Phys. Lett. B {\bf 106}, 307 (1981).
\bibitem{bh5} R. Gambini and J. Pullin,
Phys. Rev. Lett. {\bf 110}, 211301 (2013).
\bibitem{sing3} D. Garfinkle and J. M. M. Senovilla, 
Class. Quantum Grav. {\bf 32}, 124008 (2015).
\bibitem{collapse4} R. Goswami, P. S. Joshi and P. Singh, Phys. Rev. Lett. {\bf 96}, 031302 (2006).
\bibitem{sing1}  S. W. Hawking and R. Penrose, Proc. Roy. Soc. Lond. A {\bf 314}, 529 (1970). 
\bibitem{HE} S. W. Hawking and G. F. R. Ellis, {\em The Large scale structure of space-time}, Cambridge University Press (Cambridge, UK, 1973).
\bibitem{horizon1} S. A. Hayward,
Phys. Rev. D {\bf 49}, 6467 (1994).
\bibitem{bh3} S. A. Hayward, 
Phys. Rev. Lett. {\bf 96}, 031103 (2006).
\bibitem{cross3} C. Hellaby and K. Lake, Astrophys. Jour. \textbf{290}, 381 (1985).
\bibitem{cross4} C. Hellaby and K. Lake, Astrophys. Jour. \textbf{300}, 461 (1986).
\bibitem{musco} A. Helou, I. Musco and J. C. Miller Class. Quantum Grav. {\bf 34}, 135012 (2017).
\bibitem{bh4} S. Hossenfelder, L. Modesto and I. Pr\'emont-Schwarz,
Phys. Rev. D {\bf 81}, 044036 (2010).
\bibitem{matching1} W. Israel, Nuovo Cimento B {\bf 44}, 1 (1966); Nuovo Cimento B {\bf 48}, 463 (1966).
\bibitem{joshi-book} P. S. Joshi, 
{\em Gravitational Collapse and Spacetime Singularities}, 
Cambridge University Press (Cambridge, UK, 2008).
\bibitem{joshi2} P. S. Joshi and I. H. Dwivedi,
Class. Quantum Grav. {\bf 9}, L69 (1992).
\bibitem{joshi} P. S. Joshi and I. H. Dwivedi, Phys. Rev. D {\bf 47}, 5357 (1993).
\bibitem{JD} P. S. Joshi and I. H. Dwivedi,
Class. Quantum Grav. {\bf 16}, 41 (1999).
\bibitem{review} P. S. Joshi and D. Malafarina, 
Int. J. Mod. Phys. D {\bf 20}, 2641 (2011).
\bibitem{kijowski} J. Kijowski and E. Czuchry,
Class. Quantum Grav. {\bf 27}, 235007 (2010).
\bibitem{L} G. Lema\`itre, Ann. Soc. Sci. Bruxelles I, A {\bf 53}, 51 (1933); Gen. Rel. Grav. {\bf 29}, 641 (1997).
\bibitem{liu} Y. Liu, D. Malafarina, L. Modesto and C. Bambi,
Phys. Rev. D {\bf 90}, 044040 (2014). 
\bibitem{chapter} D. Malafarina, 
In: Astrophysics of Black Holes. Astrophysics and Space Science Library, Springer, Berlin, Heidelberg. Edited by C. Bambi, vol {\bf 440}, 169 (2016).
\bibitem{universe} D. Malafarina, Universe {\bf 3}, 48 (2017).
\bibitem{bobir} D. Malafarina and B. Toshmatov, 
Phys. Rev. D {\bf 105}, L121502 (2022).
\bibitem{visser-ec} P. Mart\'in-Moruno and M. Visser, 
J. High Energ. Phys. {\bf 2013}, 50 (2013).
\bibitem{mw} M. M. May, and R. H. White, 
Phys. Rev. {\bf 141}, 1232 (1966).
\bibitem{misner} C. Misner and D. Sharp, Phys. Rev. {\bf 136}, B571 (1964).
\bibitem{OV} J. R. Oppenheimer and G. M. Volkov, Phys. Rev. {\bf 56}, 374 (1939).
\bibitem{OS} J. R. Oppenheimer and H. Snyder, Phys. Rev. {\bf 56}, 455 (1939). 
\bibitem{nled} R. Pellicer and R. J. Torrence,
Journ. Math. Phys. {\bf 10}, 1718 (1969).
\bibitem{sing} R. Penrose, Phys. Rev. Lett. {\bf 14}, 57 (1965).
\bibitem{ccc} R. Penrose,
Riv. Nuovo Cim. \textbf{1}, 252 (1969).
\bibitem{peres} A. Peres,
Phys. Rev. {\bf 122}, 273 (1961).
\bibitem{poisson} E. Poisson,
{\em A Relativist's Toolkit: The Mathematics of Black-Hole Mechanics}, Cambridge University Press (Cambridge, UK, 2004). 
\bibitem{Bergmann} T. A. Roman and P. G. Bergmann, 
Phys. Rev. D {\bf 28}, 1265 (1983).
\bibitem{collapse9}
A.~Saini and D.~Stojkovic,
Phys. Rev. D \textbf{89}, 044003 (2014).
[arXiv:1401.6182 [gr-qc]].
\bibitem{collapse8} T. Schmitz,
Phys. Rev. D {\bf 103}, 064074 (2021).
\bibitem{sing2} J. M. M. Senovilla,
Gen. Rel. Gravitation {\bf 30}, 701 (1998).
\bibitem{T} R. C. Tolman, Proc. Natl. Acad. Sci. USA, {\bf 20}, 410 (1934).
\bibitem{Tooper} R. Tooper, 
Astrophys. Jour. {\bf 140}, 434 (1964).
\bibitem{bh8} B. Toshmatov, B. Ahmedov, A. Abdujabbarov and Z. Stuchl\`ik, 
Phys. Rev. D {\bf 89}, 104017 (2014).
\bibitem{nled5} B. Toshmatov, Z. Stuchl\`ik and B. Ahmedov, 
Phys. Rev. D {\bf 98}, 028501 (2018).
\bibitem{bobir2} B. Toshmatov, Z. Stuchl\`ik, B. Ahmedov and D. Malafarina,
Phys. Rev. D {\bf 99}, 064043 (2019).
\bibitem{vaidya} P. C. Vaidya, Proc. Indian Acad. Sc., {\bf A33}, 264 (1951).
\bibitem{horizon3} M. Visser, 
Phys. Rev. D {\bf 90}, 127502 (2014).
\bibitem{vaidya-gen} A. Wang and Y. Wu,
Gen. Relativ. Grav. {\bf 31}, 107 (1999).
\bibitem{will} C. M. Will, 
Living Rev. Relativity {\bf 17}, 4 (2014).
\bibitem{cross1} P. Yodzis, H.-J. Seifert and H. Muller zum Hagen, Commun. Math. Phys. \textbf{34}, 135 (1973).
\bibitem{cross2} Y. B. Zel’dovich and L. F. Grishchuk, Mon. Not. R. Astro. Soc. {\bf 2} (07), 23 (1984).
\bibitem{zen} P. Żenczykowski, 
Found. Sci. {\bf 24}, 287 (2019).









\end{thebibliography}
\end{document}